\documentclass[iop]{emulateapj}

\newcommand{\kms}{km s$^{-1}$}
\newcommand{\msun}{$M_{\sun}$}
\newcommand{\mJybeam}{mJy\thinspace beam$^{-1}$}

\newcommand{\sixcm}{$\lambda 6$ cm}
\newcommand{\HI}{\ion{H}{1}}
\newcommand{\Halpha}{H$\alpha$}
\newcommand{\HII}{\ion{H}{2}}
\newcommand{\CO}{$^{12}$CO $J=1\rightarrow 0$}
\newcommand{\Jybeamms}{Jy\thinspace beam$^{-1}$ m\thinspace s$^{-1}$}
\newcommand{\Jybeamkms}{Jy\thinspace beam$^{-1}$ km\thinspace s$^{-1}$}
\newcommand{\msunpc}{$M_{\sun}$ pc$^{-2}$}

\begin{document}


\title{Ocular Shock Front in the Colliding Galaxy IC 2163}


\author{Michele Kaufman\altaffilmark{1}, Bruce G. Elmegreen\altaffilmark{2},
Curtis Struck\altaffilmark{3}, Debra Meloy Elmegreen\altaffilmark{4}, 
Fr\'{e}d\'{e}ric Bournaud\altaffilmark{5},  Elias Brinks\altaffilmark{6}, 
Stephanie Juneau\altaffilmark{5}, and Kartik Sheth\altaffilmark{7},
}

\altaffiltext{1}{110 Westchester Rd, Newton, MA 02458, USA;  kaufmanrallis@icloud.com}

\altaffiltext{2}{IBM Research Division, T.J. Watson Research Center, 
1101 Kitchawan Rd., Yorktown Heights, NY 10598; bge@us.ibm.com}

\altaffiltext{3}{{Department of Physics \& Astronomy, Iowa State University, Ames, IA
50011; struck@iastate.edu} }

\altaffiltext{4}{Department of Physics \& Astronomy, Vassar College, Poughkeepsie, NY
12604; elmegreen@vassar.edu}

\altaffiltext{5}{Laboratoire AIM-Paris-Saclay, CEA/DSM-CNRS-Universit\'{e} Paris Diderot,
Irfu/Service d'Astrophysique, CEA Saclay, Orme des Merisiers, F-91191 Gif sur Yvette, 
France; frederic.bournaud@gmail.com; stephanie.juneau@cea.fr}

\altaffiltext{6}{{University of Hertfordshire, Centre for Astrophysics Research, College
Lane, Hatfield AL10~~9AB, United Kingdom; e.brinks@herts.ac.uk} }

\altaffiltext{7}{NASA, 300 E. Street SW, Washington, DC 20546; astrokartik@gmail.com}


\begin{abstract}

ALMA observations in the \CO\ line of the interacting galaxy pair IC 2163 and 
NGC 2207 at $2'' \times 1.5''$ resolution reveal how the encounter drives gas to pile up
in narrow, $\sim 1$ kpc wide,  ``eyelids'' in IC 2163.
 IC 2163 and NGC 2207 are involved in a grazing encounter, which has
led to the development in IC 2163 of an eye-shaped (ocular) structure at mid-radius
and two tidal arms. 
The CO data show that there are large velocity gradients across the width of each eyelid, 
with a mixture of radial 
and azimuthal streaming of gas at the outer edge of the eyelid relative to its inner edge.
The sense of the radial streaming in the eyelids is consistent with the idea that 
gas from the outer
part of IC 2163 flows inward until its radial streaming  slows down abruptly and 
the gas piles up in the eyelids. The radial compression at the eyelids causes an 
increase in the gas column density by direct radial impact and also leads to a high rate of shear. 
A linear regression of  the molecular column density $N({\rm H}_2)$ on the magnitude of 
$|dv/dR|$ across the width of the eyelid 
at fixed values of azimuth finds a strong correlation between $N({\rm H}_2)$ and $|dv/dR|$.
Substantial portions of the eyelids have high velocity dispersion in CO,  indicative of 
elevated turbulence there.

\end{abstract}


\keywords{galaxies: individual (IC 2163/NGC 2207), galaxies: ISM,
galaxies: interactions}


\section{Introduction}

The pair of galaxies \object{IC 2163} and \object{NGC 2207} 
are involved in a grazing encounter.  IC 2163 exhibits an eye-shaped (ocular) 
structure midway out in the disk and two tidal arms. In the generic N-body 
simulations of galaxy encounters  by \citet{elmegreen91},
\citet{sundin89},  and \citet{donner91}, a disk galaxy that undergoes a  close, prograde
approximately
in-plane collision with  a galaxy of similar mass develops two long tidal arms
and, for a relatively short time, an ocular structure. Although galaxy collisions of this type
are not uncommon, only a few galaxies with ocular structure are known
\citep{elmegreen95a, kaufman97, kaufman99} because the ocular stage is of short 
duration (the duration of a specific ocular disturbance is of the order of the 
ocular radius divided by the streaming motions at the ocular).  
Making detailed observations of a galaxy pair at this phase is critical for
testing numerical simulations of prograde encounters and the assumed physics.

\begin{figure*}
\epsscale{1.1}
\plotone{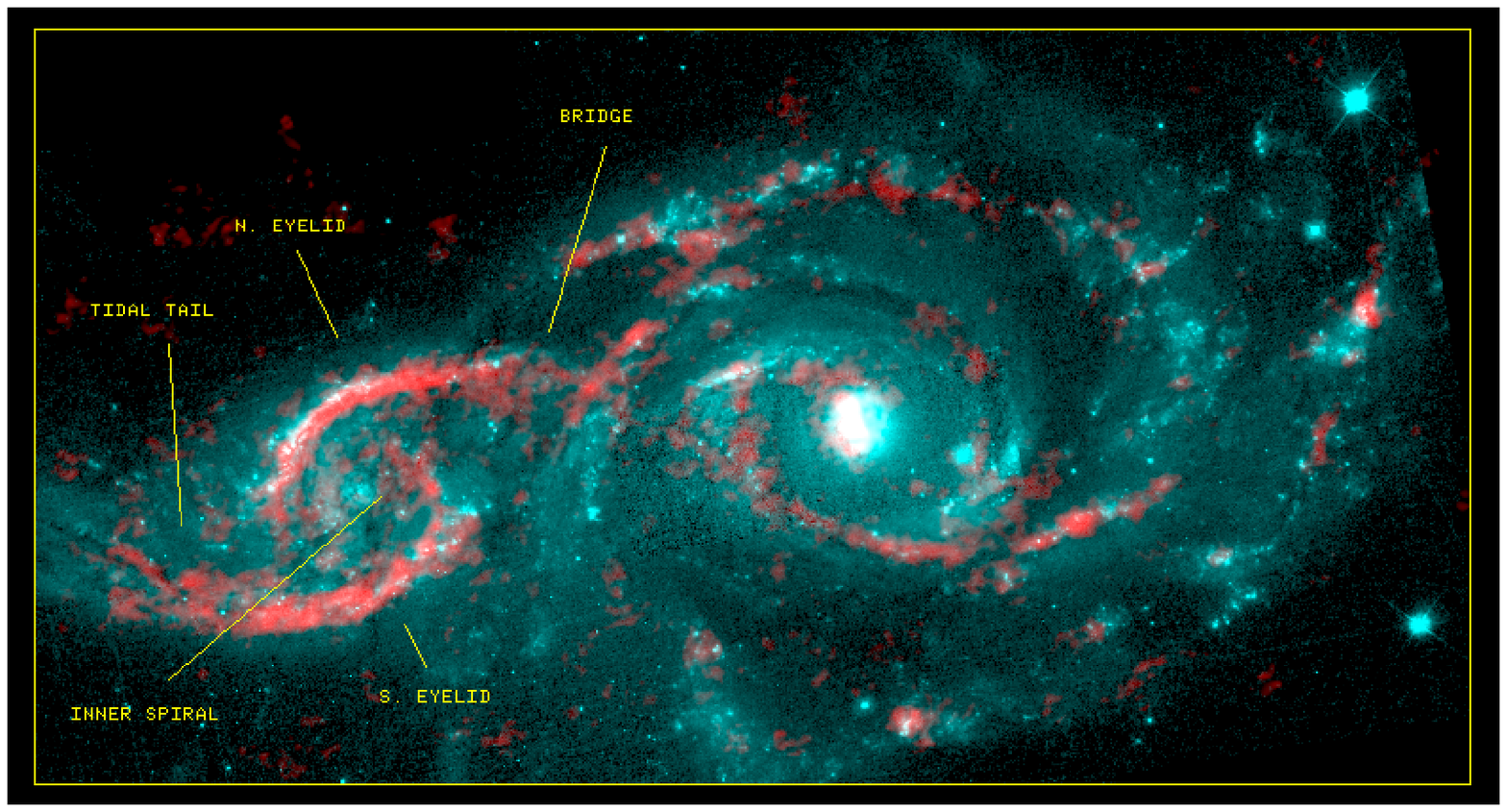}
\plotone{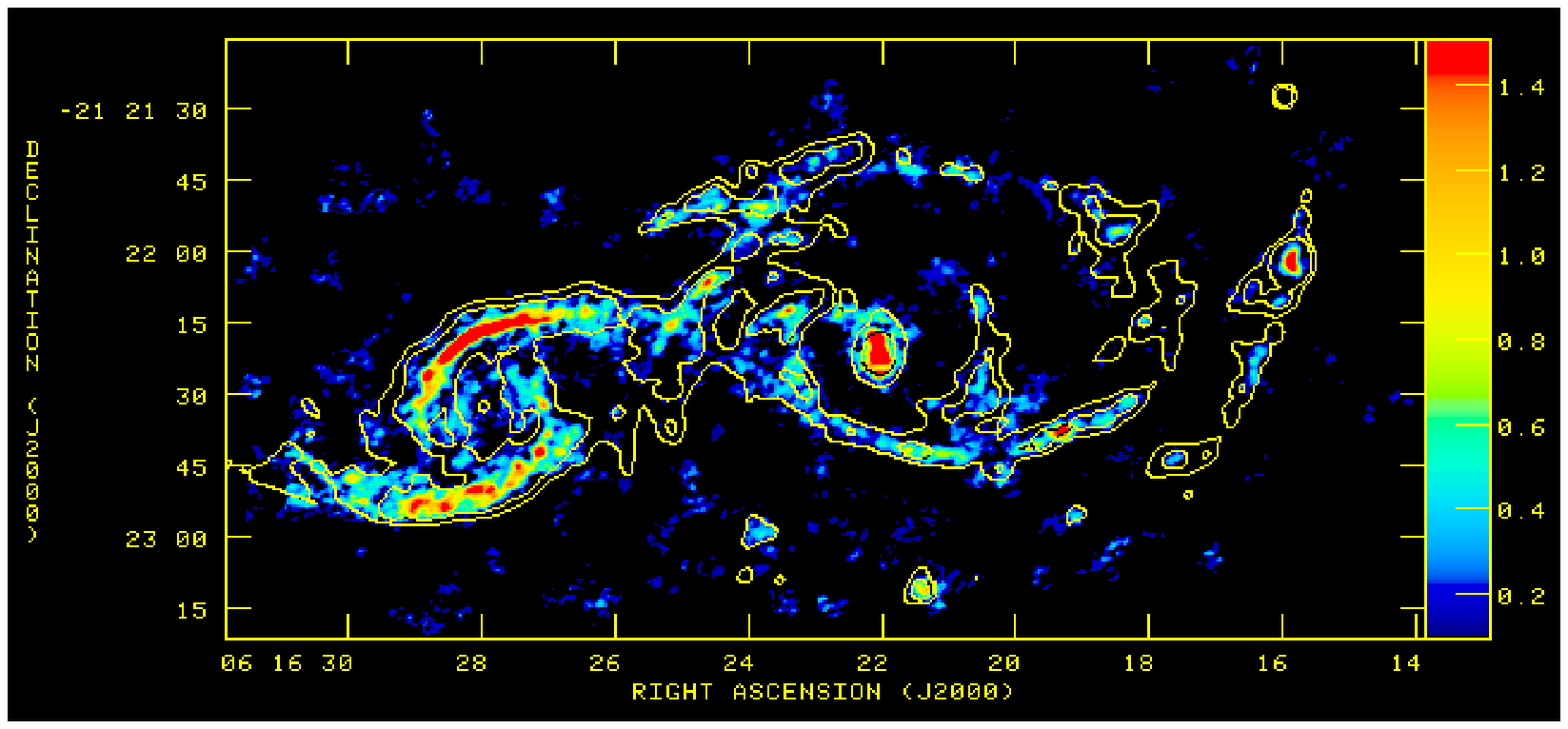}
\caption{Top: Composite color image of IC 2163 (the galaxy on the left) and
NGC 2207 (the galaxy on the right) with the {\it HST}  B image in cyan and
the \CO\ integrated intensity image in red. Bottom: Color-coded
$I$(CO) overlaid with contours of {\it Spitzer} 
8 \micron\ emission at 4 and 8 MJy sr$^{-1}$.
The wedge is in units of \Jybeamkms, where 1 Jy beam$^{-1}$
is equivalent to $T_b$ = 30.7 K, and 1 \Jybeamkms\
  corresponds to $N({\rm H}_2)$ = 88.6 \msunpc.
\label{fig1}}
\end{figure*}

IC 2163 is a prime example of an ocular galaxy. The narrow ridges along the
ocular, which we call  the  ``northern and southern eyelids''  
(labelled in the  top panel of Figure\,\ref{fig1}), 
have  a plane-of-sky width of  $\sim 6''$  as seen
in  {\it Hubble Space Telescope} WFPC2 $U B V I$ bands 
and ground-based \Halpha\  \citep{elmegreen00, elmegreen01}, 
in  {\it Spitzer} IRAC (3.6 -- 8\micron) images \citep{elmegreen06}, and  
in  Karl G.  Jansky Very Large Array (VLA)\footnote 
{The National Radio Astronomy Observatory is a facility of the National
 Science Foundation operated under cooperative agreement by Associated
 Universities, Inc.} 
$\lambda 6$ cm radio
continuum observations and {\it XMM-Newton} $UVM2$ images  \citep{kaufman12}. 
Some lower resolution
observations that also reveal the concentration of material in the eyelids are the VLA \HI\ 
and $\lambda 20$ cm radio continuum \citep{elmegreen95a}, the {\it Spitzer}  MIPS 1
(24 \micron) \citep{elmegreen06}, and the  {\it  Herschel Space Observatory} 
PACS (Photodetector
Camera and Spectrometer) 70 \micron\ maps \citep{mineo14}.

We have previously modelled NGC 2207/IC 2163, reproducing many of the observed
features with N-body \citep{elmegreen95b} and SPH
encounter simulations \citep{elmegreen00, struck05}. In the encounter simulations,
IC 2163  suffers a prograde, nearly in-plane encounter as it passes, mainly southward,
behind NGC 2207
(relative to us), and its ocular structure forms from outer disk material that loses
angular momentum in the tidal perturbation, flows inward until it hits
an angular momentum barrier,  piles up in the eyelids, and produces a shock zone there.
Although in the SPH model by \citet{struck05}, the outer disk of IC 2163 initially 
side-swipes the outer disk of NGC 2207, the ocular shock zone is not the result
of direct contact between the two galaxies but develops later from the prolonged prograde
encounter. 

\citet{elmegreen01, elmegreen06} observe bright knots of star formation along the eyelids.
\citet{kaufman12} find that the flux density ratio of 8 \micron\ to 
$\lambda 6$ cm radio continuum emission for the star-forming regions on the eyelids is a 
factor of 2 greater than elsewhere in these galaxies (or in the M81 \HII\ regions, which 
they take as a reference standard).  In that paper, we
 offer three possible explanations: shock-heated H$_2$ in the eyelids, 
shock fragmentation of large grains down to polycyclic aromatic hydrocarbon 
(PAH) sizes to increase the number of small grains, or the
accumulation of {\it B} stars in the eyelids from the previous 30 Myr of star formation.

We observed IC 2163/NGC 2207 in \CO\  at  a resolution of
$2'' \times 1.5''$ with the Atacama Large Millimeter Array (ALMA).
The goal of the present paper is to
use the ALMA data to look for the predicted gas flows into the eyelids that drive 
the pileup of gas there
 and to relate the observed streaming motions to the molecular gas densities in
the ocular shock zone.  This is a major test of the numerical simulations,
and our results provide constraints on future models. 
 In the  \HI\ observations by \citet{elmegreen95a} there are 
indications of streaming motions around the oval. However the $13.5'' \times 12''$ resolution
of the \HI\ data is too low compared to the $\sim 6''$ width of the eyelid to get
detailed information. In contrast, the spatial
resolution afforded by the ALMA CO  observations allows us to distinguish between
the outer and the inner edges of each eyelid and thus  measure the radial and tangential streaming of the gas at the outer edge of an eyelid relative to that at its inner edge.

From the NASA/IPAC Extragalactic  Database (NED),
we adopt a distance of $35 \pm 2.5 $ Mpc for IC 2163/NGC 2207 with a
Hubble constant $H$ = 73 \kms\ Mpc$^{-1}$ and 
corrected for infall towards Virgo. Then $1''$ =  170 pc, and the $\sim 6''$ 
width of the eyelids corresponds to 1 kpc.

The velocities listed in this paper are heliocentric and, unless stated otherwise, use the 
radio definition of the nonrelativistic Doppler shift.

Section 2 describes our ALMA observations.
Section 3 measures the velocity differences $\Delta v$ and $dv/dR$
across the width of each eyelid and compares the \HI\ and CO velocity fields at the
ocular. Section 4 presents the correlation  between $dv/dR$
and the molecular column densities $N({\rm H}_2)$  in the eyelids.  
Section 5 considers evidence for turbulence in the eyelids,
 and  Section 6 summarizes our conclusions.

\section{\label{observe} Observations and data reduction}

On 2014 April 3, IC 2163 and NGC 2207 were observed in \CO\ emission
in a mosaic of 34 pointings
 with the ALMA 12-m array of 32 antennae (two of which were subsequently completely flagged
 throughout the run due to outlying values of $T_{\rm sys}$).
 The observations are from Configuration C32-4 of Cycle 1.
The on-source time was 38 minutes and the pointings were
$26^{\prime\prime}$ apart. The phase center was R.A.,  decl. (2000) = 06 16 22.809, -21 22
30.71. Ganymede served as the flux calibrator with the Butler-JPL-Horizons 2012
frequency-dependent flux model,  J0609-1542 served as the bandpass calibrator, and
 J0609-1542 served as the phase calibrator. Hanning smoothing was applied.

The uv-coverage ranged from 8.0 to 169 k$\lambda$. 
The maximum recoverable scale with this ALMA configuration is $15''$ = 2.6 kpc,
which is appreciably broader than the IC 2163 eyelids or the spiral arms of NGC 2207. 
The only single-dish \CO\ observations of this galaxy pair are a shallow mapping using
several pointings with the Swedish ESO  Submillimeter Telescope (SEST)
(HPBW = $43''$)  by \citet{thomasson04}, who does not list the
 total integrated CO flux.  Due to lack of access to any suitable observations, our 
interferometric data could not be combined with single-dish observations. 
As described in \citet{struck05}, 
the  SEST observations find CO emission from both disks, with brighter emission from IC 2163,
and the CO emission does not seem to be concentrated on the massive
\HI\ clouds identified by \citet{elmegreen95a}.  The SEST spectrum of IC 2163 is 
double peaked, hinting that some of the emission is from the eyelids.
The missing short-spacing data in our
ALMA maps could lead to underestimates of  the total CO flux of  each galaxy. However,
because diffuse CO emission will be 
more spread out than our 2.6 kpc maximum recoverable scale, it 
should make only a minimal contribution to the molecular column densities of the
clumped emission we are studying and is unlikely to affect the velocities at the 
eyelids which are the main focus of this paper.

The CASA 4.2.1 software package was used for the data reduction. Calibration and flagging 
were carried out by the ALMA Data  Reduction Team. We subsequently checked the data and
flagged one high amplitude spike in the target observations. 
Using the CASA clean routine, a cleaned data cube of line emission from the galaxy pair
was made with natural weighting, a central frequency of 114.1973 GHz and a total bandwidth
of 0.3845 GHz.  

For the data analysis we used the AIPS software package.
To select areas of genuine CO emission, we convolved the above cube to $6''$
resolution, clipped it at 2.5 times its rms noise, and retained regions of emission only
if they appear in at least two adjacent channels.
The result was applied as a blanking
mask to the original cube, and after correcting  for primary-beam attentuation, we took
 a $551 \times 301$ pixel $\times 64$ channel subcube which contains all of
the line-emission in the masked, primary-beam corrected cube.  

Table\,\ref{OBS} lists the properties of 
this subcube,  which was used for calculating the moment maps. 
We also blanked the image of
the intensity-weighted velocity field and  
the velocity dispersion image
where the CO surface brightness $I{\rm (CO)} \leq 200$ \Jybeamms\
in the CO surface density map.
A surface brightness of 200 \Jybeamms\ is equivalent to
$2.7 \times$  the rms noise over two channel widths.

\begin{deluxetable}{lc}
\tablewidth{0pt}
\tablecaption{Final ALMA \CO\ Subcube\tablenotemark{a}
\label{OBS}}
\tablehead{
    \colhead{Parameter}  &  \colhead{Value} }
\startdata
Configuration               &     C32-4 \\
Channel width               &    10 \kms\   \\
 Pixel size                      &    $0.5''$    \\
 Cube Size                     &    $551 \times 301$ pixels,  64 channels \\
 Weighting                    &    Natural  \\
 PSF (HPBW, PA)             &  $2.00'' \times 1.52'' $,  $-68.5\degr$  \\
 rms noise per channel  &   3.7 mJy beam$^{-1}$ \\
 $T_b/I$(CO)                 &   30.7 K/Jy beam$^{-1}$  \\
Velocity range of line-flux  & $2517 -   3067$  \kms\    \\
Total integrated $S$(CO)  &   $500 \pm 5$  Jy \kms\  \\
Peak Dynamic range\tablenotemark{b}   
                                     &  24  \\
\enddata 
\tablenotetext{a} {Velocities are heliocentric, radio definition}\
\tablenotetext{b} {Peak brightness in the channel maps is on the northern eyelid.}
\end{deluxetable}  

Except as noted in Section 3.1 below, the correction for primary beam attenuation is a factor of 
$1.0 -  1.2$  for almost all of the  CO emission from the galaxy pair in the
ALMA map.

To convert the values of CO surface brightness to molecular column density, we 
take $X_{CO}$ = $1.8 \pm 0.3 \times 10^{20}$ (H$_2$ cm$^{-2}$)/(K \kms) from
\citet{dame01}.

The observations also yielded a continuum image centered on 102.3 GHz. The brightest
continuum source is well below the rms noise in the spectral-line cube.

We compare our CO images of IC 2163  with the VLA \HI\ observations by \citet{elmegreen95a}.
The latter have a point spread function (PSF) 
with FWHM equal to $13.5'' \times 12''$, PA = $90\degr$.
We use an \HI\ surface density image 
  made from a masked cube with a channel width of 21 \kms\ and
rms noise per channel = 0.73 mJy beam$^{-1}$ 
and  an \HI\ velocity field image made from a masked
cube with a channel width of 5.25 \kms\ and rms noise = 1.3 mJy beam$^{-1}$.

We also display the HST image from \citet{elmegreen00}, the \Halpha\ image from
\citet{elmegreen01},  the {\it Spitzer} 8 \micron\ image 
from \citet{elmegreen06}, and the  HiRes deconvolution image from \cite{velusamy08} of the 
{\it Spitzer} 24 \micron\ data.

\section{\label{vel} Velocities at the Eyelids}

\subsection{CO Velocity Differences across the Eyelid Width}

The top panel in Figure\,\ref{fig1} displays a  composite color image of
IC 2163/NGC 2207 with the {\it HST} {\it B} image from \citet{elmegreen00} in cyan and the 
surface brightness image from our ALMA CO data in red.
Features of interest in IC 2163 are marked in this figure. In the bottom panel of
Figure\,\ref{fig1} the  color-coded $I$(CO) image is overlaid
with contours of {\it Spitzer}  8 \micron\ emission at 4 and 8 MJy sr$^{-1}$.
 This figure  shows that along the eyelids in IC 2163 and along the spiral arms of NGC 2207, the
CO emission occupies the same narrow ridge as the 8 \micron\ emission.  For clarity,
higher level and lower level contours of 8 \micron\ surface brightness are omitted. 
The molecular gas in IC 2163 is strongly concentrated in the northern and southern eyelids,
with fainter CO emission from the inner spiral arms, the portion of the tidal tail included
in this field, and part of the tidal bridge extension of the northen eyelid, until it blends,
in projection, with a spiral arm of NGC 2207.  The eyelids are generally much
brighter in CO than the spiral arms of NGC 2207; the highest molecular
column density in the galaxy pair (including the nucleus of NGC 2207)
is on the northern eyelid.

\begin{figure*}
\epsscale{1.1}
\plotone{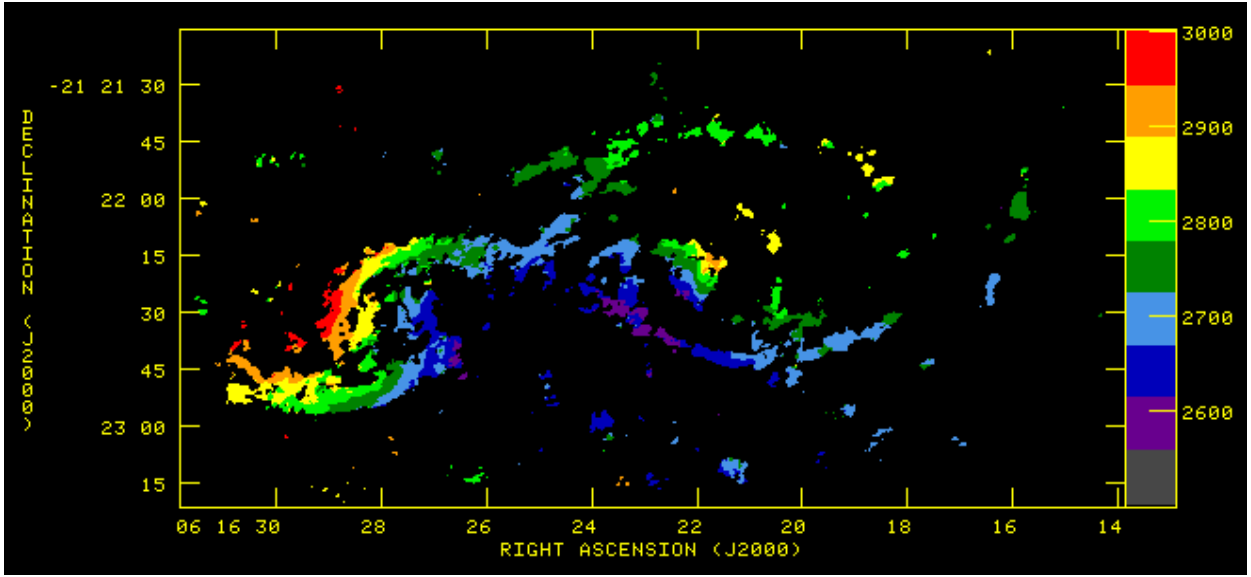}
\caption{CO velocity field of IC 2163 and NGC 2207. The wedge is in units of \kms. There is a
large difference in line-of-sight velocity across the $\sim 6''$ width of each eyelid.
\label{fig2}}
\end{figure*}

In the outer northeastern part of the field in Figure\,\ref{fig1}
 (R.A.  east of 06 16 28 with decl. north of
-21 21 50), no \HI\ emission is detected [see Section 3.2 below and \citet{elmegreen16}],
so the CO emission there is probably not real. At this outlying location and
at an analogous  outlying location in the southwestern part of the field in Figure\,\ref{fig1},
the correction for CO primary beam exceeds a factor of 1.3.

Figure\,\ref{fig2} displays the CO velocity field  of IC 2163/NGC 2207. 
We note that the CO isovelocity contours
on the eyelids make a shallow angle with respect  to the outer edge of each eyelid. 
The velocity field image reveals that over a wide range of azimuth, there is a large velocity 
gradient across the $\sim 1$
kpc width of the eyelid at any given azimuth.  We define the line-of-sight velocity 
difference between the outer and the inner edge of the eyelid as
\begin{equation}
\Delta v \equiv v_{\rm outer} - v_{\rm inner}.
\end{equation}

The values of $\Delta v$  for the eyelids  are much greater than the velocity differences
 across the width of the spiral arms in NGC 2207.  At some locations along the eyelids,
cuts across the eyelid width  perpendicular to the eyelid ridge-line yield values of $\Delta v$ in 
excess of 100 \kms.

\begin{figure}
\epsscale{1.1}
\plotone{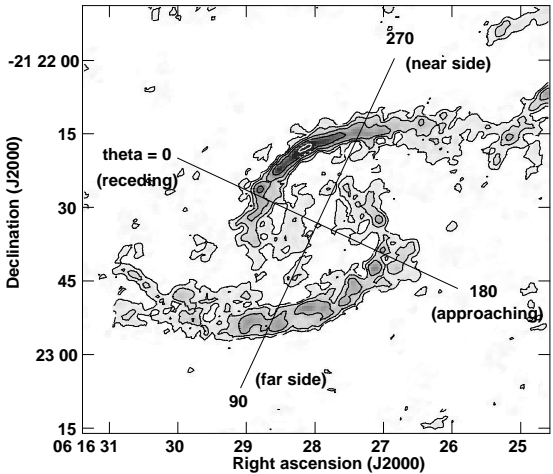}
\caption{Grayscale and contour display of  $I$(CO) image of  IC 2163.  Contour levels
are at 200, 500, 1000, 2000, and 5000 \Jybeamms, where 1 Jy beam$^{-1}$
is equivalent to $T_b$ = 30.7 K and the lowest contour level is equivalent to
$2.7 \times$  the rms noise over two channel widths. 
The long lines labelled with the values of the azimuth $\theta$ mark the kinematic
major axis at $\theta$ = $0\degr$ and $180\degr$ and the \HI\ kinematic minor axis  at $\theta$ =
$90\degr$ and $270\degr$, with $\theta$ measured ccw from the receding major axis.
The brighest CO emission in the galaxy pair occurs on the northern eyelid 
and has a maximum $I$(CO)
 of $5.64 \times 10^3$ \Jybeamms,  corresponding to  an LOS
$N({\rm H}_2)$ of  500 \msun\ pc$^{-2}$.
\label{fig3}}
\end{figure}

For  measurement and analysis of the velocities at the eyelids,
 we transformed the images of IC 2163 to face-on polar coordinates.  We adopted 
values of the projection parameters of IC 2163 from \citet{elmegreen95a, elmegreen95b},
i.e., we take the intersection of the plane of the disk with the plane of sky on the receding side as
the \HI\ kinematic major axis at position angle PA = $65\degr$ and the disk 
inclination $i$  as $40\degr $ (where $i$ = 0 for face-on).
 The ocular is intrinsically more oval than it appears in the sky-plane
\citep{elmegreen95b}. For the transformation to polar coordinates, we take the center as
the nucleus, which is at the center of the oval. The receding, approaching, near, and far sides 
of IC 2163 are identified in the sky-plane image of the CO surface brightness in 
Figure\,\ref{fig3}. The long lines in this figure are labelled with the value of the azimuthal angle 
$\theta$ in face-on polar coordinates, i.e.,  the kinematic major axis at $\theta$ =
$0\degr$ and $180\degr$ and the kinematic minor axis at $\theta$ =
$90\degr$ and $270\degr$.
The azimuth $\theta$ is measured counterclockwise 
from the receding kinematic major axis at PA = $65\degr$.

\begin{figure*}
\epsscale{1.1}
\plottwo{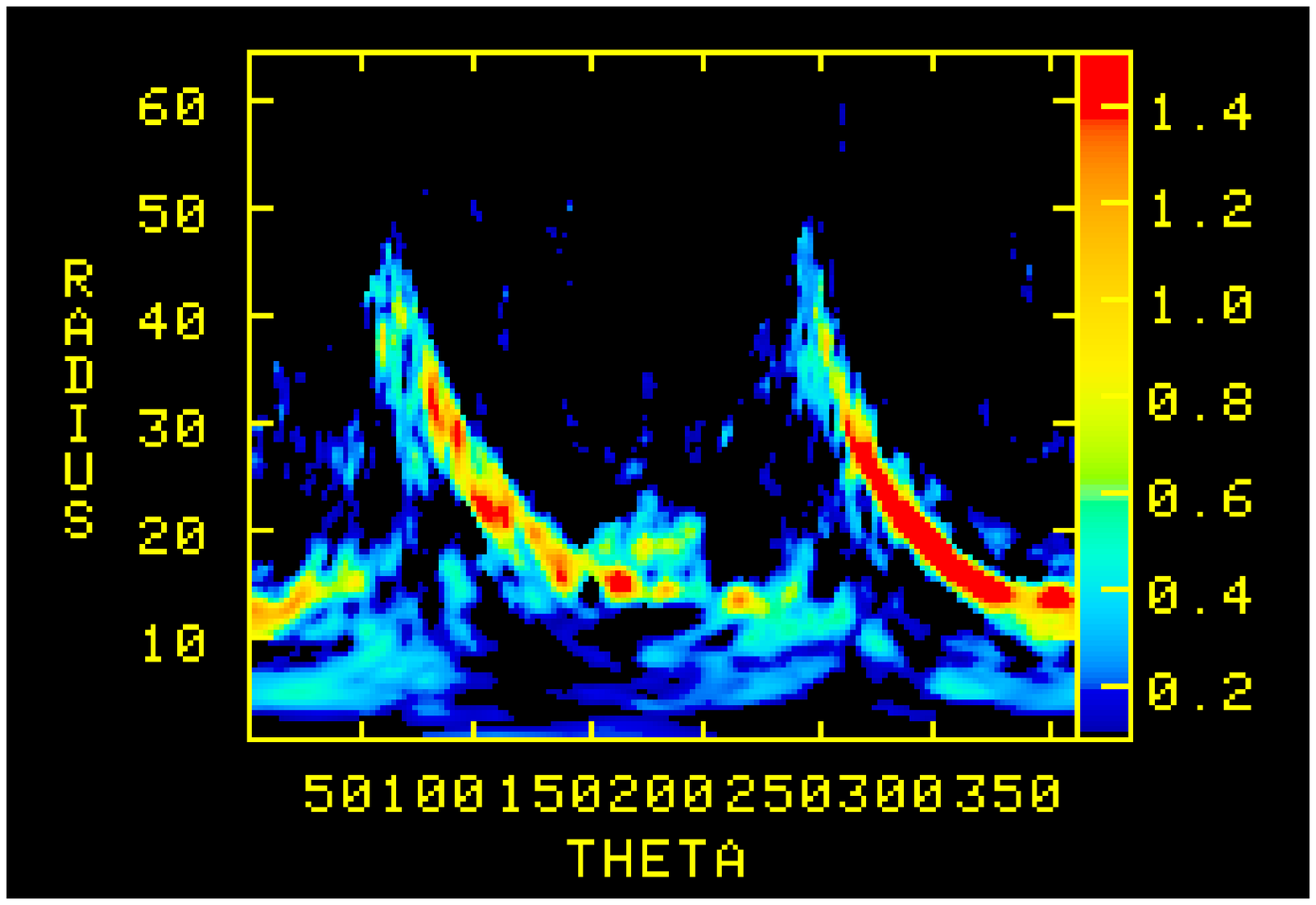}{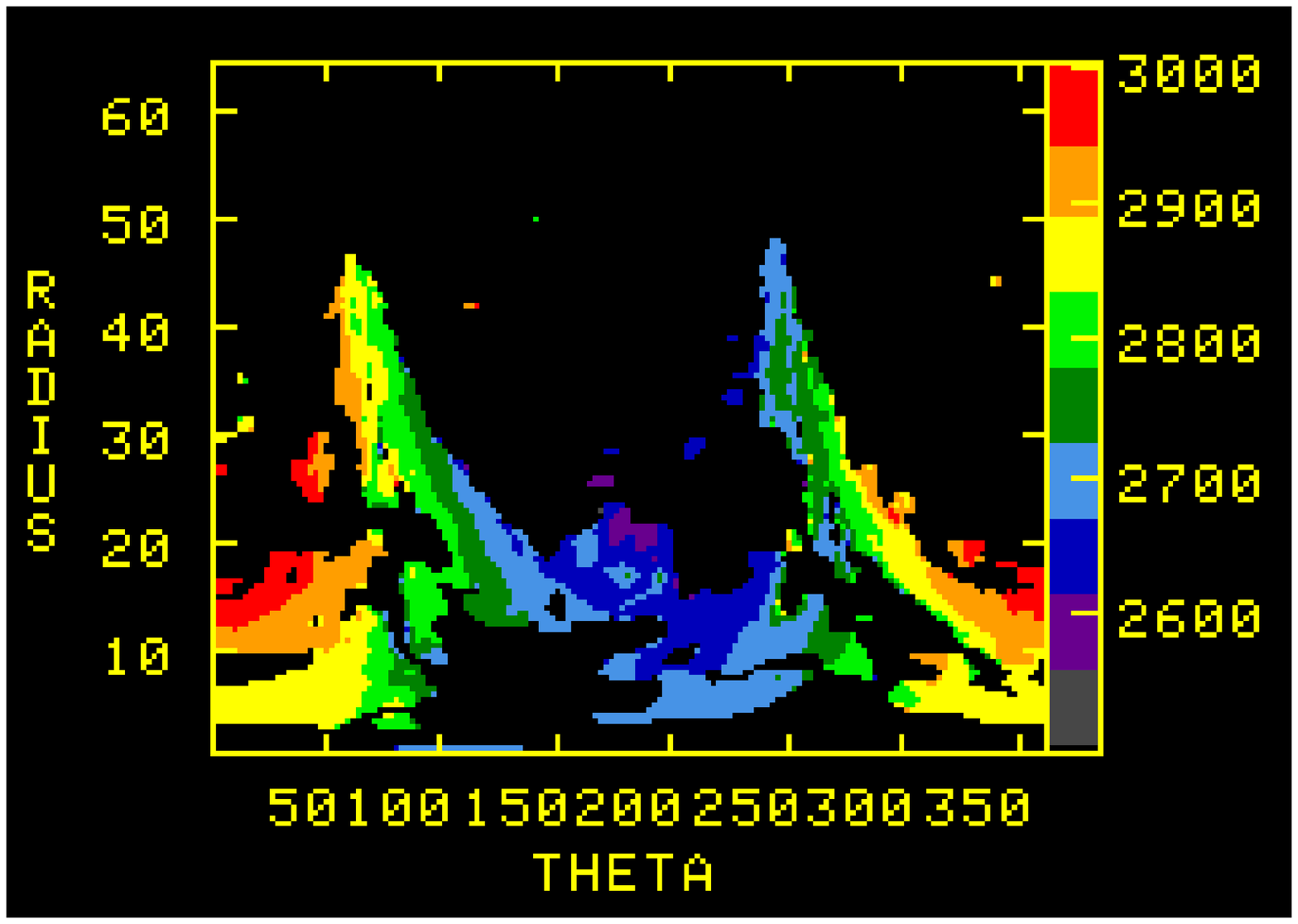}
\caption{Images in face-on polar coordinates, with the CO integrated intensity image on
the left and the CO velocity field on the right. The units of the $I$(CO) wedge are 
  \Jybeamkms,  and the units of the velocity wedge are \kms. The azimuth $\theta$ is measured 
ccw from the receding kinematic major axis. The radius is in arcsec.
\label{fig4}}
\end{figure*}

Figure\,\ref{fig4} displays the CO surface brightness image and the CO velocity field 
in face-on polar coordinates.  The \HI\ kinematic minor axis on the side nearest us 
 is at $\theta = 270\degr$ (as indicated  in Figure\,\ref{fig3}).
Then the observed line-of-sight velocity
  \begin{eqnarray}
  v_{\rm obs}(R,\theta) = v_{\rm sys} + [v_R(R,\theta) \sin\theta \nonumber\\
+v_\theta(R,\theta) \cos\theta] \sin i + v_z \cos i,
  \end{eqnarray}
where $R$ is the face-on galactocentric radius, $v_R$ is positive for expansion, and $v_\theta$ is
the sum of the circular velocity $v_c$ and tangential streaming $v_t$. Because the tidal
peturbation to IC 2163 is nearly in-plane, we assume that across the eyelid width,
 $\Delta v_z$ for the molecular gas is negligible compared to $\Delta v_R$  or $\Delta v_t$.
 In Figure\,\ref{fig4}, the southern eyelid 
starts at $\theta \approx 75\degr$, and  the northern
eyelid starts at $\theta \approx 250\degr$.

For the eyelid measurements 
of each parameter, i.e., the line-of-sight (LOS) velocity,  
the face-on radius $R$, the difference in face-on radius 
$\Delta R$ = $R_{\rm outer} - R_{\rm inner}$ 
between the outer and inner edges of the eyelid, and the LOS column density
$N({\rm H}_2)$,  we defined the edge of the eyelid
as the closest pixel position to where $I$(CO) =  200 \Jybeamms\ except for the following
two transition regions at the ends of the eyelids. Specifically, to distinguish between CO emission
from the southern eyelid and that from the inner spiral at $\theta = 167\degr - 174\degr$
and to distinguish  between CO emission from the northern eyelid and that from the
transition to the tidal tail at $\theta = 28\degr - 37\degr$, we used the location of 
IRAC 4 clumps as a guide.
The transformation from plane-of-sky to face-on polar coordinates resulted in a pixel size 
of $2.3\degr$  for the azimuthal angle $\theta$.  We measured
the values of $\Delta v$ at one pixel ($2.3\degr$) intervals of $\theta$ in the polar coordinate image, 
with the line-of-sight velocity measured at the same value of
$\theta$ at both edges. We then averaged our measurements over four pixels 
($9.3\degr$) in $\theta$ as that gives an arclength $R \Delta \theta$ approximately equal to the
FWHM of the PSF at the smallest value of  $R$ on the inner edge of an eyelid.

Table\,\ref{table2}
lists the values of the LOS velocity difference
$\Delta v$ and the difference in face-on radius $\Delta R$ 
between the outer and inner edges of the eyelid.
The values of  $N({\rm H}_2)$  in this table
  are averages for the area between the inner and outer edges of  the eyelid.
For $\Delta R$ and $N({\rm H}_2)$, we used the same values of
 $\theta$  and  averaged over 
the same $9.3\degr$ in azimuth as for $\Delta v$.
Given the channel width, the minimum uncertainty in $\Delta v$ is 7 \kms. 
The uncertainties listed for $\Delta v$ in Table\,\ref{table2}  are either 7  \kms\  or the
standard error of the mean in averaging over $9.3\degr$ in $\theta$, whichever
is greater.  The rms noise in the CO channel maps times the channel width is 
37 \Jybeamms, equivalent to $N({\rm H}_2)$ = 3.3 \msunpc. The uncertainties listed for
$N({\rm H}_2)$ in Table\,\ref{table2} are either 3.3 \msunpc\ or the standard error of
the mean in averaging over $9.3\degr$ in $\theta$, whichever is larger.

The observed large values of $\Delta v$ over a wide range of azimuth (see Table\,\ref{table2})
argue for
a mixture of radial and azimuthal streaming of gas at
 the outer edge of the eyelid with respect to  gas  at the inner edge of the eyelid.
With the adopted orientation of IC 2163,
outer disk gas that streams radially inward, slows down radially in the ocular 
ridge and speeds up azimuthally there,
 would produce $\Delta v$ positive on much of the 
 northern eyelid,  i.e., from $\theta = 270\degr - 360\degr$,
 and $\Delta v$ negative  on much of  the southern eyelid, i.e., from $\theta = 90\degr - 180\degr$. 
This is in the same sense as the observed values. 
Thus the observed CO velocity field is 
consistent with the idea that gas from the outer part of IC 2163 flows inward until its
radial streaming slows down abruptly  to produce the thin pileup zone in the eyelids. 

\begin{deluxetable*}{ccccc}
\tablewidth{0pt}
\tablecaption{Eyelid Velocity Differences and $N({\rm H}_2)$\tablenotemark{a,b}
\label{table2}}
\tablehead{
  \colhead{$\theta$}  & \colhead{$\Delta v$} & \colhead{$N({\rm H}_2)$}  & \colhead{$\Delta R$} 
   & \colhead{$dv/dR $}  \\
  \colhead{($\degr$)} & \colhead{(\kms)}       & \colhead{(\msunpc)} & \colhead{(arcsec)} 
  &  \colhead{(\kms\ kpc$^{-1}$)}  \\
  \colhead{(1)}  & \colhead{(2)}  & \colhead{(3)}  & \colhead{(4)}  &  \colhead{(5)}   \\
 }
\startdata
Northern Eyelid&&&\\  
 &\\     
      263.6             & $ 31 \pm 7$           &  $ 63.9  \pm 4.9$          & 10.4    &  $ 27 \pm 6$   \\ 
      272.9             & $110 \pm 26$        &  $ 91.7  \pm 3.4$          &   8.1    & $124  \pm 29$ \\ 
      282.2             & $127 \pm 22$        &  $114.6 \pm 3.3$          &   7.2    & $161 \pm 28$  \\ 
      291.5             & $179 \pm 20$        &  $153.3 \pm 16.1$        &   6.5    & $252 \pm 28$  \\              
      300.8             & $148 \pm 28$        &  $170.8  \pm 5.6$         &   7.3    & $186 \pm 35$  \\
      310.1             & $102 \pm 24$        &  $151.3 \pm 3.8$          &   4.5    & $207 \pm 49$  \\ 
      319.4             & $117 \pm 12$        &  $142.9 \pm 4.3$          &   4.2    & $255 \pm 26$  \\  
      328.7             &   $71 \pm 7$          &  $113.9 \pm 14.2$        &   4.3    & $151 \pm 15$  \\
      338.0             &   $40  \pm 9$         &   $ 71.9 \pm 3.3$          &   4.5    & $ 81 \pm 18$   \\   
      347.2             &   $58 \pm 8$          &   $ 84.5 \pm 4.7$          &   5.6    & $ 95 \pm 13$   \\ 
      356.5             &   $63 \pm 7$          &   $ 76.4 \pm 4.0$          &  7.4     & $ 78 \pm 9$   \\ 
          5.8             &   $61 \pm 7$          &   $ 63.8 \pm 3.3$          &  5.8     & $ 96  \pm 11$  \\
        15.1             &   $57 \pm 7$          &   $ 62.0 \pm 3.3$          &  4.0     & $130 \pm 16$  \\ 
        24.4             &   $48 \pm 7$          &   $ 65.7 \pm 3.3$          &  4.6     & $ 95  \pm 14$  \\ 
        33.7             &   $37 \pm 7$          &   $ 48.6 \pm 4.0$          &  3.8     & $ 89  \pm 17$ \\ 
&\\
Southern Eyelid&&&&\\
&\\
        75.5            &  $    9 \pm 20$       &   $ 40.3 \pm 7.0$           & 10.0    &   $8  \pm 18$  \\
        84.8            &  $-59 \pm 14$       &   $ 84.5 \pm 6.0$           &  8.8     &  $- 61 \pm 14$ \\
        94.1            &  $-59 \pm  7$        &   $ 86.3 \pm 6.7$           &  6.5     &  $- 83 \pm 10$  \\    
       103.4           &  $-60 \pm 11$       &   $ 86.4 \pm 5.0$           &  6.7     &   $-82 \pm 15$ \\     
       112.7           &  $-63 \pm 7$         &   $ 90.1 \pm 4.4$           &  6.5     &   $-89 \pm 10$ \\    
       122.0           &  $-69 \pm 7$         &   $ 54.0 \pm 3.6$           &  6.9     &   $-92 \pm 9$  \\   
       131.2           &  $-57 \pm 7$         &   $ 67.7 \pm 3.3$           &  6.5     &  $- 80 \pm 10$  \\   
       140.5           &  $-55 \pm 7$         &   $ 80.3 \pm 6.0$          &  5.1     &  $-99  \pm 13$   \\     
       151.1           & \nodata                  &  \nodata        & \nodata  & \nodata  \\
       161.4           &  $-28 \pm 7$         &   $105.1 \pm 5.6$          &  3.0     &   $-85 \pm 21$  \\     
       170.7           &   $  2  \pm 10$       &   $ 65.8 \pm 9.0$           &  3.5     &   $5 \pm 25$  \\    
\enddata
\tablenotetext{a} {From measurements made at fixed values of $\theta$ in
face-on polar coordinate images and averaged over $9.3\degr$ in  $\theta$. 
The azimuth $\theta$ is measured ccw from the receding kinematic major axis
at PA = $65\degr$. }
\tablenotetext{b} {$\Delta v$ is the LOS velocity difference between the outer and inner
edges of the eyelid, $N({\rm H}_2)$ is the LOS molecular column density averaged over the 
area between the same outer and inner edges for the same fixed azimuth as $\Delta v$,
$\Delta R$ is the difference in face-on radius,  and $dv/dR$
 is the difference approximation to the directional derivative in the disk-plane
 of $v$ with respect to $R$ for $\theta$ fixed and inclination $i = 40\degr$.
 }
\end{deluxetable*}

In Table\,\ref{table2} the positive values of $\Delta v$ for $\theta = 6\degr - 34\degr$,
as the gas in the northern eyelid  approaches the \HI\ tidal tail,
probably mean that the $\Delta v_\theta \cos \theta$ term in Equation 2  dominates in
this range.
The values of $|\Delta v|$ are smaller at the extreme eastern and extreme western ends 
of each eyelid than elsewhere on the eyelid; according to the encounter
simulations, the material there
originated in regions  that experienced weaker tidal torque.

On the northern eyelid at $\theta = 270\degr$, 
 $\Delta v$ results from radial streaming only and has an LOS
value of $110 \pm 26$ \kms, equivalent to $\Delta v/\sin i$ =$171 \pm 40$ \kms\ in the plane
of the disk. 
At  $\theta =0\degr$ on the northern eyelid, $\Delta v$ results from azimuthal motions
only, and the value of 
$\Delta v/\sin i$ is $96 \pm 11$ \kms.  Thus there is considerable shear associated with tangential
streaming since 
it is unlikely that the circular velocity changes by
an amount this large across the 1 kpc width of the eyelid.  
The tangential streaming
at $\theta = 0\degr$ has an  appreciably  smaller magnitude than the
radial streaming at $\theta = 270\degr$, but  the values of $v_R$ and $v_t$ 
probably vary with $\theta$. 
On the southern eyelid at $\theta = 90\degr$, $|\Delta v|$ results from radial streaming only,
and the value of $|\Delta v|/\sin i$ = $ 92 \pm 11$ \kms. 

Column (5) of Table\,\ref{table2} lists the difference approximation to the directional derivative in the
disk-plane of the velocity
 with respect to $R$ across the width of the  eyelids for $\theta$ fixed:
\begin{equation}
     \frac{dv}{dR} = \frac{\Delta v (\sin i)^{-1}}{\Delta R},
\end{equation}
 where  $\Delta v$ is the LOS velocity difference from  the Column (2) 
 and $\Delta R$ is the difference in face-on radius in kpc.  
Note that this includes radial and tangential streaming.  
Except at the extreme eastern
and extreme western ends of the southern eyelid and at the extreme western end of
the northern eyelid, the values of $|dv/dR|$ are all quite large, with values in excess of
150 \kms\ kpc$^{-1}$ on a significant part of the northern eyelid and 
$80 - 90$ \kms\ kpc$^{-1}$ on the southern eyelid.
The velocity gradient
is greater than or equal to the directional derivative, and thus the velocity gradients across the
eyelids are large. Such large values are qualitatively consistent with the encounter simulation
predictions that  the eyelid compression region is a shock zone. 

For the rest of this paper, we shall 
assume that the tangential streaming component $\Delta v_t$ dominates $\Delta v_c$ at
the eyelids.

If we take the duration of a specific ocular disturbance to be of the order of the 
ocular radius divided by the streaming motions at the ocular, then the present ocular
structure in IC 2163 should last for a timescale of the order of several $\times 10^7$ yr.

\begin{figure}
\epsscale{1.0}
\plotone{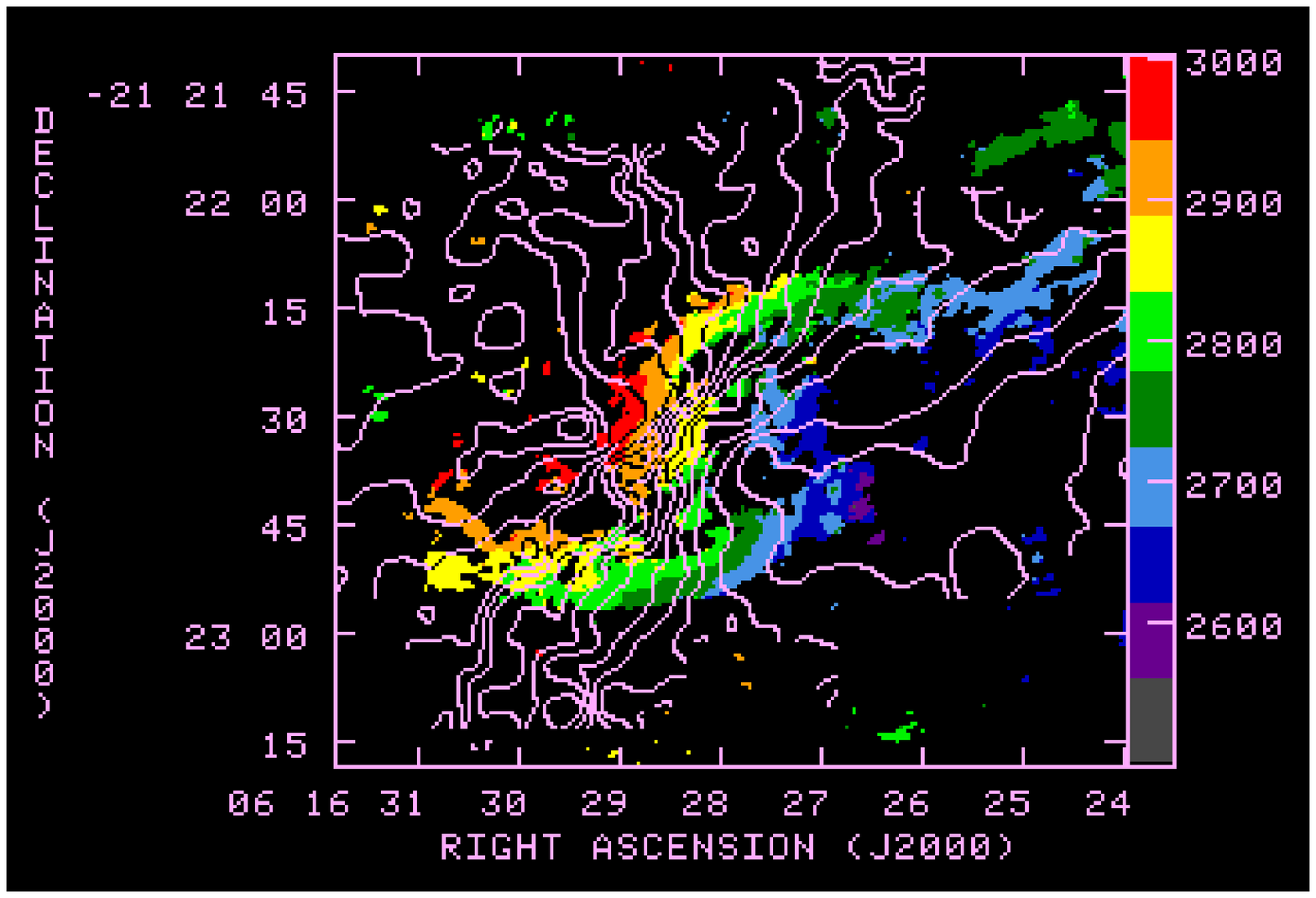}
\plotone{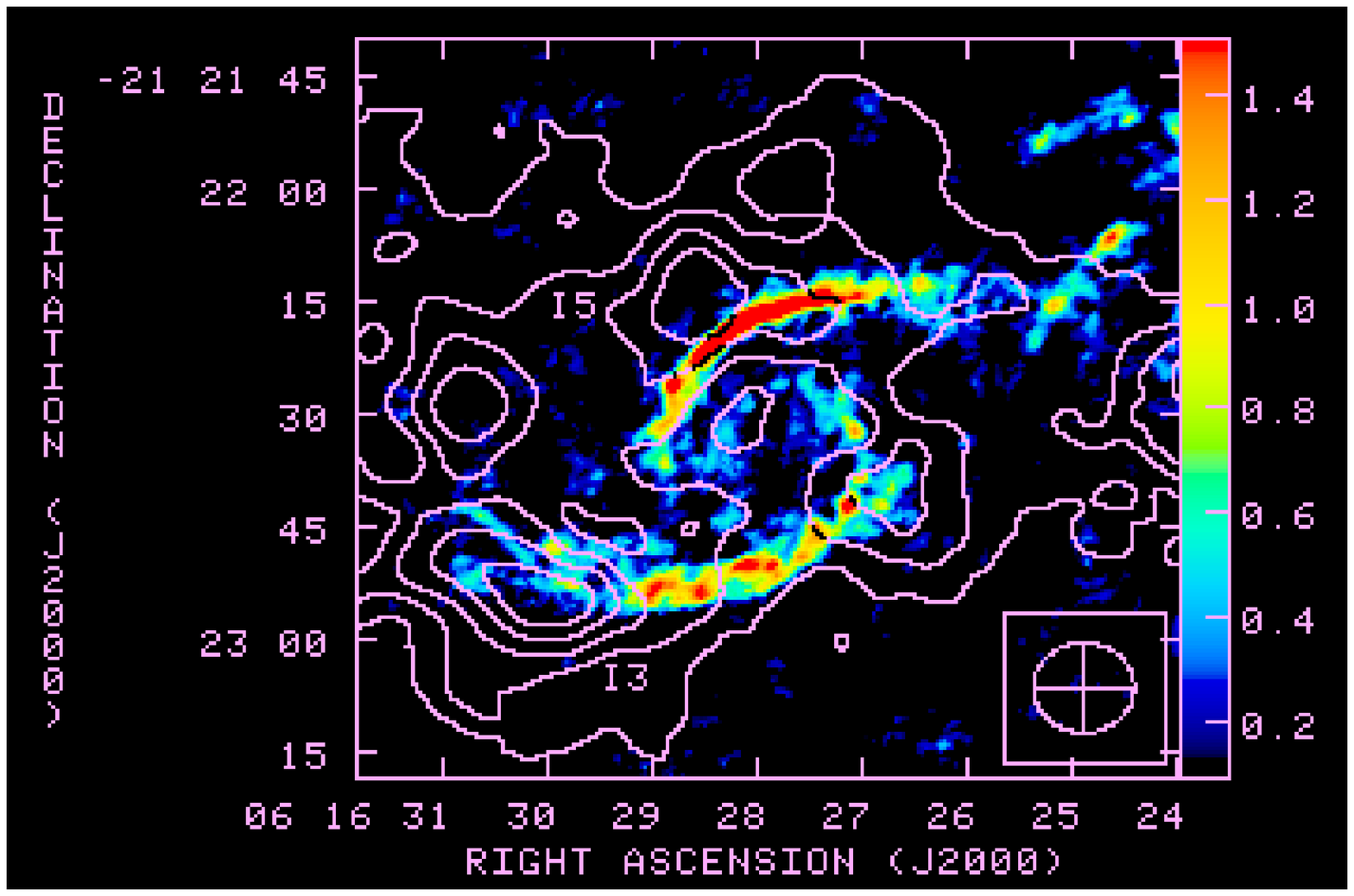}
\caption{Top: \HI\ isovelocity contours of IC 2163 overlaid on the CO velocity field. The
CO velocity wedge is in units of \kms, and
 the \HI\ velocity contours are at 2640 ... (20) ... 2960  \kms.
Bottom: $N$(\HI) contours of IC 2163 overlaid on $I$(CO).
 The \HI\ contours levels are  5, 10, 15, 20, and 25 \msunpc.  The $I$(CO)
 wedge is in units of \Jybeamkms,  where 1 Jy beam$^{-1}$
 is equivalent to $T_b$ = 30.7 K, and 1 \Jybeamkms\ corresponds to a molecular column density of 
88.6 \msunpc. The lowest $N$(\HI) contour level is equivalent to 
$3 \times$ the rms noise over two channel  widths. The beam symbol represents the
\HI\ resolution. Massive \HI\ clouds I3 and I5 from \citet{elmegreen95a} are labelled.
\label{fig5}}
\end{figure}

\subsection{Comparison between CO and \HI\ Velocity Fields}

\begin{figure}
\epsscale{1.0}
\plotone{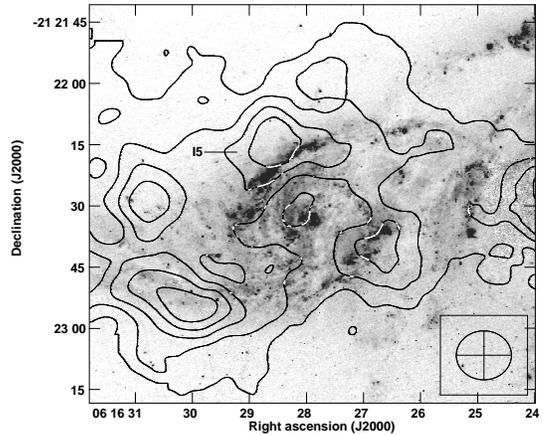}
\caption{$N$(\HI) contours of IC 2163 from the bottom panel of Figure\,\ref{fig5} 
overlaid on the {\it HST} B image in
grayscale. The beam symbol represents the \HI\ resolution. The massive \HI\ cloud I5 is
labelled.
\label{fig6}}
\end{figure}

\citet{elmegreen95a}  separate the \HI\ associated with IC 2163 from
that associated with NGC 2207. 
  For IC 2163, the top panel of  Figure\,\ref{fig5} displays \HI\ isovelocity
contours overlaid on the CO velocity field, the bottom panel of Figure\,\ref{fig5}
 has \HI\ column densiy contours overlaid on $I$(CO),
 and Figure\,\ref{fig6}  has the same $N$(\HI) contours
overlaid  on the {\it HST} B-band image.  The lowest $N$(\HI) contour level in these figures 
 is 5 \msunpc, which is equivalent to $3 \times$ the rms noise over
two channel widths. The beam symbol 
 represents the \HI\ resolution; the \HI\ point spread function (PSF)  
has FWHM equal to $13.5'' \times 12''$, PA = $90\degr$. 
Thus the \HI\ images  provide a considerably less detailed view  than our CO data
with a resolution of $2'' \times 1.5''$.

 In the CO velocity  
field, the kinematic major axis is at  position angle PA =  $65\degr \pm 5\degr$:
on the outer edge of the eyelids, 
 the highest velocity molecular gas is at PA = $65\degr$ on the northern eyelid and
diametrically opposite the lowest velocity molecular gas on the southern eyelid. 
This is consistent with the value  PA =  $65\degr \pm 10\degr$
that \citet{elmegreen95a} obtain for the kinematic major axis in \HI.

Because of tidal effects, the velocity fields in \HI\ and CO are not regular, and  in
\HI\ the dynamic center appears to be $\sim 5''$ northeast of the nucleus.
The largest amplitude wiggles in the \HI\ isovelocity contours tend to be displaced toward the 
outer side of  the CO eyelids.  Not all of the wiggles in \HI\ velocity contours just north of the 
northern eyelid have the same sense of curvature; in the vicinity of the massive \HI\ cloud I5 
\citep{elmegreen95a}, labelled in the bottom panel of Figure\,\ref{fig5},
 the  curvature of the wiggles in the \HI\ velocity contours 
reverses sense. This indicates complex flows in the \HI\ gas related to this
massive \HI\ cloud. There are also pronounced wiggles in the \HI\ 
velocity contours near the massive \HI\ cloud I3, which is centered at the start of the tidal tail.

At the CO eyelids,  the CO and \HI\ isovelocity 
contours have a similar tilt at some locations but not  at other locations, particularly
on the southern eyelid. 
Some differences between the tilt of the CO and \HI\
isovelocity contours at the eyelids may result from differences in spatial resolution and  
differences in where the column densities of molecular and atomic gas have local maxima. 

The \HI\ velocity dispersion at the massive \HI\ clouds I5 and I3 is $\sim 50$ \kms\ after correction
for the velocity gradient across the $\sim 2$ kpc synthesized beam \citep{elmegreen95a}.
\citet{elmegreen00} suggest that
these high values for the \HI\ velocity dispersion could result from  (a) high turbulence 
in the \HI\ gas and lead to  unusually large \HI\  scale heights 
or  from (b) several streams with different velocities within the  \HI\ synthesized beam. 
If the \HI\ disk is thicker than normal, the \HI\ velocities at higher altitudes may differ 
from those in the disk.
For comparison, Section 5 below presents our measured values of the CO velocity dispersion
in various parts of the galaxy pair.

The massive \HI\ cloud I5,  
whose center lies outside of  the CO eyelid and outside of  the B-band 
ridge,  is located northeast of the 
region of greatest $N({\rm H}_2)$ on the northern eyelid. 
Based on \Halpha\ and 24 \micron\ emission, \citet{elmegreen16}  find a very low star formation
rate in their aperture A10, which covers most of  I5.
Thus the \HI\ here is not a product of photodissociation.  I5 may be along (or part of) the stream
producing the greatest concentration of molecular gas on the eyelids, and some of  its \HI\ 
may get converted into molecular gas in the eyelid compression zone.

\subsection{Differences between North and South Eyelids}

We compare the CO velocity field values of 
$|\Delta v|/(\sin i) $  and  $|dv/dR|$ on the northern eyelid with those on
the southern eyelid. 

For $\theta = 273\degr  - 310\degr $ on the northern eyelid,
the streaming motions in the disk $\Delta v/(\sin i) $ range in magnitude from
  159 \kms\ to  278  \kms, and  the values of $dv/dR$ are $124 - 252$ 
  \kms\ kpc$^{-1}$.  Since this range of $\theta$ lies within
$40\degr$ of  the  \HI\
kinematic minor axis, these velocities represent mainly radially inward  motion of
molecular gas at the outer edge of the northern eyelid relative to its inner edge.
On the southern eyelid
for $\theta = 94\degr - 131\degr$, there is little
variation of $\Delta v$  or of $dv/dR$  with azimuth; the values of
$\Delta v/(\sin i) $ here are  negative and have a mean magnitude $93 \pm 11$ \kms, 
and $dv/dR$ has a mean magnitude of $85 \pm 11$ \kms\ kpc$^{-1}$.
Since this range of $\theta$ is within $41\degr$
 of the kinematic minor axis, the observed negative values of $dv/dR$ on the southern 
eyelid here represent mainly 
radially  inward motion of molecular gas at the outer edge of the southern eyelid
relative to its inner edge. 
   Thus in the CO velocity field the magnitude of
the radial streaming motions  and the values of $|dv/dR|$ near the \HI\ kinematic minor axis are
considerably higher on the northern eyelid and vary much more with azimuth $\theta$
than on the southern eyelid.

Our measurements find a relative velocity difference between the outer and inner edges
of each eyelid in the direction that produces a compression zone at the eyelids.
Although the sense of this velocity difference is consistent with having gas from 
the outer part of the galaxy flow inwards until it reaches the eyelids, an alternative
interpretation would have the inner edge of the eyelids moving outwards with respect to
the outer edge. To see if the inner edge of the eyelids is moving radially relative to the
center of the disk, we compare the CO value of $v_{\rm inner}$ (the velocity at the inner 
edge of the eyelid) on the kinematic minor axis with $v_{sys}$.
The values of $v_{sys}$ (heliocentric, optical definition) listed by  \citet{elmegreen95a} 
are $2775 \pm 10$ \kms\ from the kinematic minor axis of their \HI\ velocity field
and $2756 \pm 15$ \kms\ from optical observations at the nucleus.  
The CO emission at the IC 2163 nucleus is faint and thus not useful for 
measuring $v_{sys}$.

After our CO velocities 
are converted to heliocentric, optical definition, the northern eyelid has
$v_{\rm inner}$ = $2777  \pm 25$ \kms\ at $\theta = 270\degr \pm 5\degr$,
consistent with the \HI\ and optical values of $v_{sys}$.
 However the southern eyelid has  $v_{\rm inner}$ = $2842 \pm 9$ \kms\
at $\theta = 90\degr \pm 3\degr$, which is appreciably higher 
than $v_{sys}$. The difference between the values of $v_{\rm inner}$ 
on the northern and southern eyelids at the kinematic minor axis
can be seen in the color displays of the CO velocity field in Figure\,\ref{fig2} and 
Figure\,\ref{fig5} where the northern eyelid is dark green and the southern eyelid is light green 
at these positions.
Relative to the center of the disk, the inner edge of the
northern eyelid does not seem to be moving radially  
but the inner edge of the southern eyelid is moving radially outwards.
After the gas streamed radially inward to form the eyelids, the inner edge of the
southern eyelid may have started to move outward in a second ocular wave.  Under
certain conditions in the models by \citet{struck05},  a second ocular wave can form and 
propagate outward.

As noted in Section 3.2 above,  the \HI\ velocity field reveals a small apparent shift of the 
dynamic center of IC 2163 to the northeast of the center of the galaxy.

These various asymmetries 
 may be evidence of a dynamic situation in which the evolution of the ocular is not symmetric about
the center of the galaxy because of  differences in
the near-field tidal forces on the two eyelids as discussed in Section 3.4 below.

\subsection{A Possible Qualitative Explanation for the Velocity Asymmetries}

Section 3.3 points out certain velocity asymmetries in the ocular. 
Relative to us, the near side of IC 2163 is at PA = $335\degr$. According to the model by 
\citet{elmegreen95b}, this is also the side of IC 2163 that was  closest to NGC 2207 
as IC 2163 traveled behind NGC 2207 from the north and west to its present position
partially behind the outer eastern side of NGC 2207. This means that most of the northern eyelid,
in particular the part with values of $\Delta v >100$ \kms,
is now on the side closest to NGC 2207. Thus the velocity asymmetries 
 may result from differences in the near-field tidal forces on the two eyelids. 
The western end of the northern
eyelid connects smoothly in shape and velocity to the tidal bridge arm between the galaxies.
The eastern end of the
southern eyelid connects smoothly in shape and velocity  to the shocked
gas returning to the main disk from the southern side (leading edge) of the broad \HI\  tidal tail
(see Figure\,\ref{fig1} and the CO and \HI\ velocity fields in Figure\,\ref{fig5}), which 
 is on the side of IC 2163 farthest from NGC 2207.

\begin{figure}
\epsscale{1.2}
\plotone{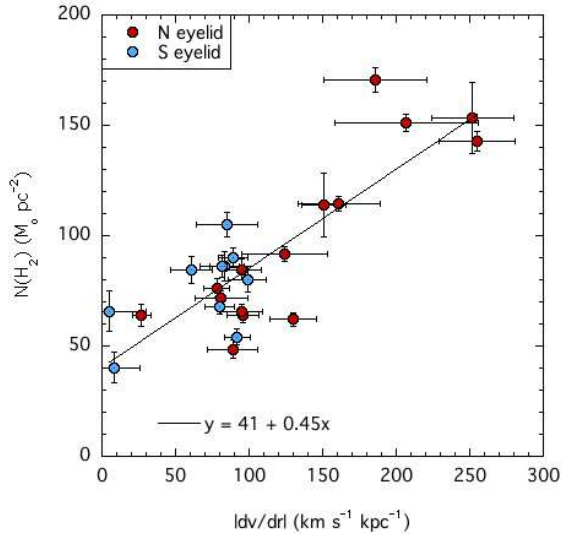}
\caption{$N({\rm H}_2)$ versus $|dv/dR|$ for 15 values of the azimuth $\theta$
 on the northern eyelid (the red points) and for 10 values of $\theta$  on the 
southern eyelid (the blue points).  The line
is the fit to all 25 points and has a correlation coefficient of
0.84.  
\label{fig7}}
\end{figure}

\section{Correlation between $|dv/dR|$ and the Molecular Column Densities}

\subsection{Observed Correlation}

 We consider what effect the large differences in velocity across the $\sim 1$ kpc width
of each eyelid  have had on the molecular column densities there.
In Figure\,\ref{fig7}, we plot the LOS molecular column density $N({\rm H}_2)$ from
Column (3) of Table\,\ref{table2} versus the magnitude of the directional derivative 
$|dv/dR|$  
of the velocity with respect to R  from Column (5) of that table.
The measurements for the 15 values of azimuth $\theta$ on the northern 
eyelid are  plotted as red points; those for the 10 values of $\theta$ on the southern eyelid
are plotted as blue points. 
 A linear regression of $N({\rm H}_2)$ on $|dv/dR|$ for all 25 points in this figure gives
\begin{equation}
  N({\rm H}_2) = 41 + (0.45 \pm 0.06) \vert\frac{dv}{dR} \vert ,
\end{equation}
for $N({\rm H}_2)$ in \msunpc\ and $dv/dR$ in \kms\ kpc$^{-1}$.
The correlation coefficient is 0.84 with 23 degrees of freedom, and
thus there is a strong correlation. Because the northern eyelid has  
a substantially larger range of $|dv/dR|$ than the southern eyelid, the 
regression tends to be dominated by the data on the northern eyelid. 
 A linear regression for the 15 values of
$\theta$  on the northern eyelid (the red points in Figure\,\ref{fig7}) gives 
\begin{equation}
  N({\rm H}_2) = 28 + (0.52 \pm 0.08) \vert\frac{dv}{dR} \vert 
\end{equation}
and a correlation coefficient of 0.87 with 13 degrees of freedom.

The $|dv/dR|$ values in this figure involve a combination of radial and tangential streaming.
We assume that at the eyelids  the tangential streaming $\Delta v_t \simeq \Delta v_\theta$.
For the northern eyelid, the point for $\theta \simeq 270\degr$ (where 
the measured $|dv/dR|$ represents pure radial streaming)
and the point for $\theta \simeq 0\degr$ (where the measured $|dv/dR|$
represents pure tangential streaming) lie on or very
close to the fitted line in Figure\,\ref{fig7}. 
We  do not measure the density contrast 
(eyelid/pre-eyelid) directly since our interferometric observations do not detect the 
pre-eyelid molecular column density. 

Our choice of value for $X_{\rm CO}$ is listed in Section 2. If we were to choose a different
value of $X_{\rm CO}$ for the eyelids, without varying its value with azimuth along 
the eyelid, this would change the slope of the fitted line in Figure\,\ref{fig7}, but
there would still be a strong correlation between $N({\rm H}_2)$ and $|dv/dR|$.

The following factors may lead to some of the scatter seen in Figure\,\ref{fig7} and/or decrease
the slope of the line.
(a)  After losing some azimuthal  velocity in the shock/shear zone
in the eyelids, the gas will start to fall inward slowly. This would tend to decrease its
density at the inner edge of the eyelid and should have a stronger effect for
stronger shocks. In Figure \,\ref{fig4} the CO emission from the southern eyelid appears 
more ratty at the inner edge than at the outer edge, which may be a consequence of 
shocks generating turbulence.
(b) According to the encounter simulations, the gas reaching the eyelid comes in from a range of radii $R$.

\subsection{Interpretation of Correlation between $|dv/dR|$ and the Molecular Column Densities}

The radial compression at the eyelids causes an increase in the gas column density by
direct radial impact and also leads to a high rate of shear there. 
One expects shocks and compression from the radial streaming and also from the shear
associated with tangential streaming; both types of streaming can lead to intersecting gas streams.

The following argument based on angular momentum conservation computes  the rate of
shear arising from radial streaming into the eyelids.
Consider particles in an annulus bounded by radii $R_2$ and $R_1$  with $R_1 < R_2$
and initial azimuthal velocities $v_{\theta,2}$ = $v_{\theta,1}$ = $v_\theta^*$ (for 
a flat rotation curve). If a perturbation kicks particles inward from $R_2$ to radius
$R_{\rm outer}$ (partway between $R_1$ and $R_2$)  and kicks particles outward from
$R_1$ to radius $R_{\rm inner} $ to form a thinner annulus bounded by $R_{\rm inner}$
and $R_{\rm outer}$, with $R_{\rm inner} <  R_{\rm outer}$, then by angular momentum
conservation, the azimuthal velocity of particles  initially at $R_2$ speeds up to
  \begin{equation}
     v_{\theta,\rm{outer}} = v_\theta^*( \frac{R_2}{R_{\rm outer}} ),
  \end{equation}
and the azimuthal velocity of particles initially at $R_1$ slows down to
    \begin{equation}
     v_{\theta,\rm{inner}} = v_\theta^*( \frac{R_1}{R_{\rm inner}} ).
  \end{equation}
The rate of shear between $R_{\rm outer}$ and $R_{\rm inner}$ 
  \begin{equation}
   \gamma = \frac{v_{\theta,\rm{outer}} - v_{\theta,\rm{inner}} } {R_{\rm outer} - R_{\rm inner}}.
  \end{equation}
Substituting for $v_{\theta,\rm{outer}}$ and $v_{\theta,\rm{inner}}$ gives
  \begin{eqnarray}
  \gamma = v_\theta^*\frac{(R_2/R_{\rm outer} - R_1/R_{\rm inner})} {R_{\rm outer} - R_{\rm inner}}
\nonumber\\
  =   v_\theta^*\frac{(R_2 R_{\rm inner} - R_1 R_{\rm outer})} {R_{\rm outer} R_{\rm inner} 
 (R_{\rm outer} - R_{\rm inner})}.
  \end{eqnarray}
Then setting $R_{\rm outer} \approx R_{\rm inner} \approx R$, this reduces to
   \begin{equation}
     \gamma = (v_\theta^*/R) \frac{R_2 -R_1}{ R_{\rm outer} - R_{\rm inner}}.
   \end{equation}
The density contrast is
   \begin{equation}
   \frac{\rho_{\rm compressed}}{\rho_{\rm initial}} = \frac{R_2 -R_1}{ R_{\rm outer} - R_{\rm inner}}.
\end{equation}
Comparison of Equations (10) and (11) shows the proportionality between the rate of shear 
 and the density contrast:
   \begin{equation}
  \frac{\rho_{\rm compressed}}{\rho_{\rm initial}} =  (R/v_\theta^* )\gamma,
   \end{equation}
where $\gamma$ = $\Delta v_\theta/\Delta R$ is the rate of shear.

The rate of shear and the column density contrast correlate with each other because
both result from the radial compression of the eyelid.

From our data, we are unable to 
disentangle how much of the observed correlation  in Figure\,\ref{fig7} is
due to direct radial impact and how much is a subsequent consequence of the 
shocks in the intersecting gas streams that arise from shear.

\begin{figure*}
\epsscale{1.0}
\plotone{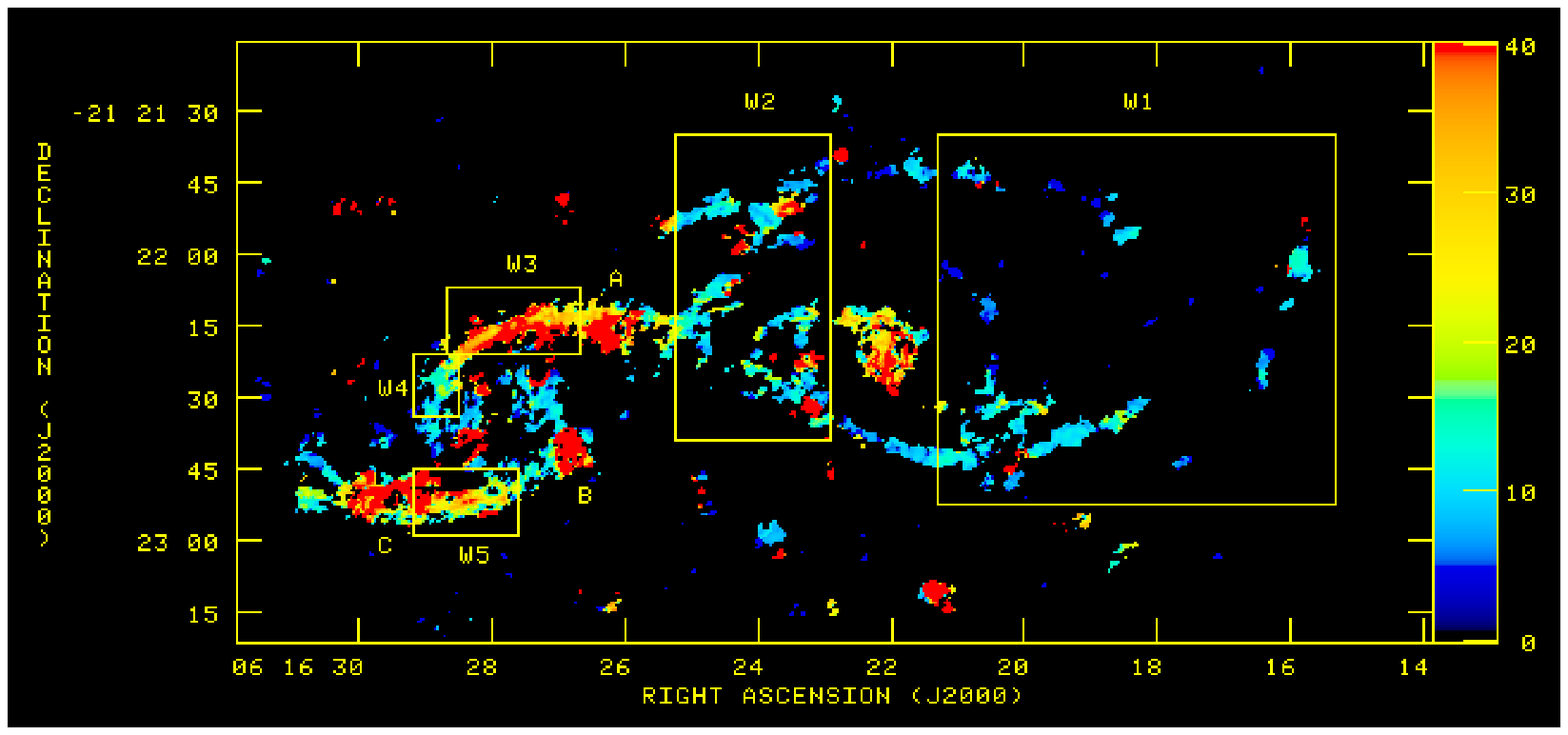}
\plotone{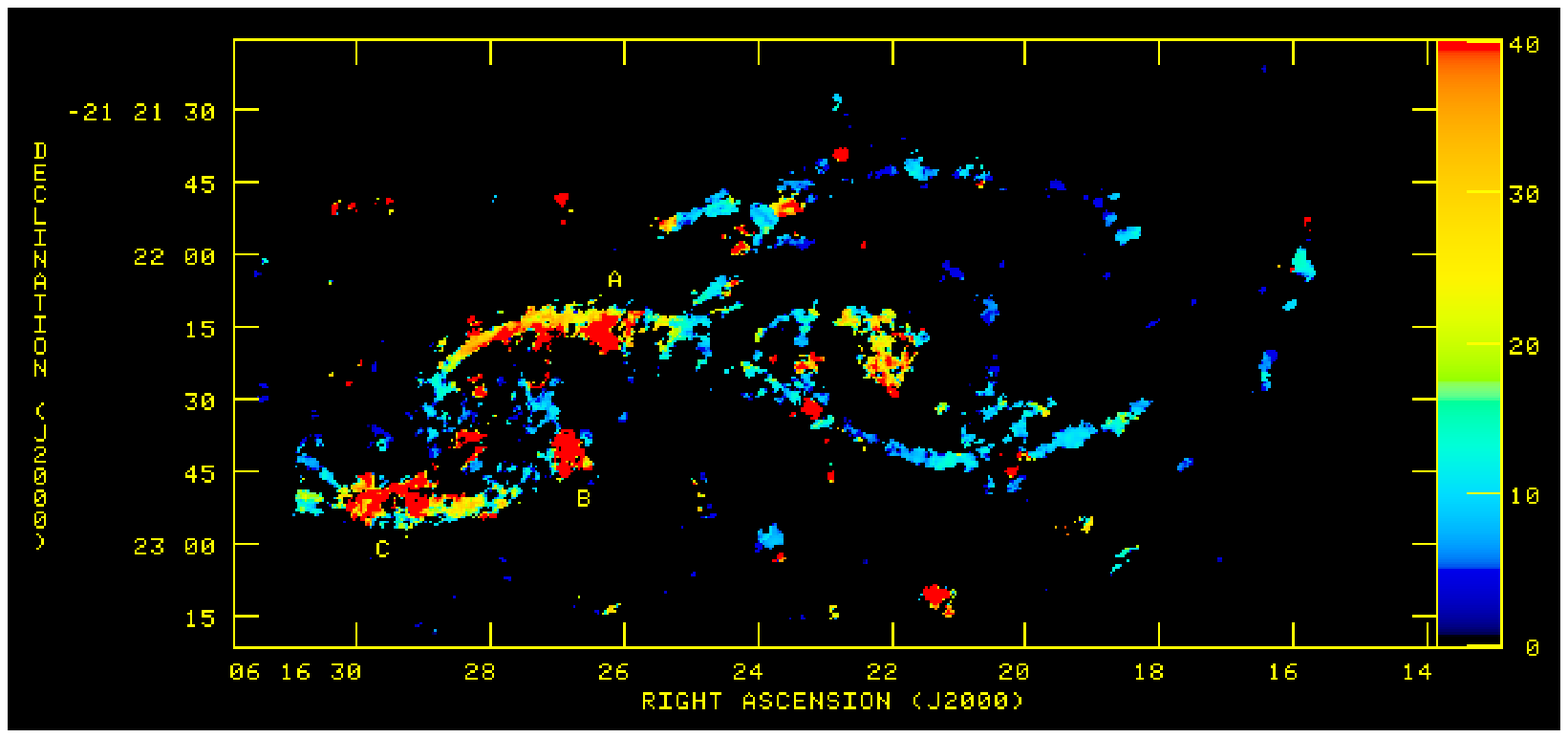}
\caption{Top: CO velocity dispersion image  with wedge in units of \kms\ 
before correction for velocity gradient across the
FWHM of the PSF.  Table\,\ref{dispersion} lists the mean values of the velocity dispersion for
the boxes labelled with W's. Bottom: CO velocity dispersion 
 after correction for  the  velocity gradient.  The velocity dispersion on a substantial 
part of the eyelids of IC 2163 is significantly greater than in the spiral arms of NGC 2207.  See text
about the dynamically complex regions labelled A, B, and C.  
\label{fig8}}
\end{figure*}

\begin{figure*}
\epsscale{0.9}
\plottwo{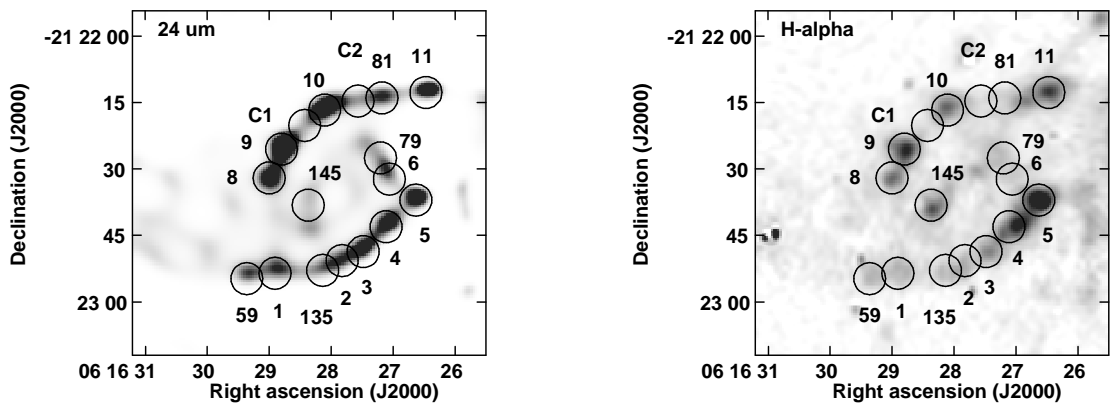}{figure9b}
\caption{Left: HiRes 24 \micron\ image of IC2163. Right: \Halpha\ image of IC 2163. 
Numbered circles ($3.6''$ in radius) mark the locations of the brighter {\it Spitzer}
8 \micron\ star-forming clumps listed by \citet{elmegreen06},
except for the two circles labelled C1 and C2, which are discussed in
the text.
\label{fig9}}
\end{figure*}

\section{CO Velocity Dispersion}

Since intersecting gas streams and shocks in the eyelid compression zone should increase
the turbulence there, we consider whether the CO observations show evidence of  high
velocity dispersion in the eyelids. 

Figure\,\ref{fig8} displays the CO velocity dispersion images before and after correcting 
 for the mean velocity gradient across the FWHM of the
PSF.  For the various portions of the galaxy pair indicated by the boxes in the top panel
of this figure, Table\,\ref{dispersion} lists the mean values of the uncorrected CO velocity dispersion 
$\sigma_v$,  the corrected velocity dispersion $(\sigma_v)_{\rm corr}$, and their rms scatter.

For the spiral arms of NGC 2207 on its western (anti-companion) side, the mean 
 $(\sigma_v)_{\rm corr}$ $\pm$ rms scatter is $10 \pm 10$ \kms.
This is consistent with the value of the CO velocity dispersion that  \citet{mogotsi16} find for
bright CO emission in normal, nearby galaxies; their average for the CO mean velocity dispersions
in 11 galaxies is $10.5 \pm 3.6$ \kms, with a range of 7 to 15 \kms.
Our mean value of $(\sigma_v)_{\rm corr}$ for the spiral arms of NGC 2207 on its companion side is 
somewhat higher than on its western side.

\begin{deluxetable}{lll}
\tablewidth{0pt}
\tablecaption{CO Velocity Dispersion
\label{dispersion}}
\tablehead{
    \colhead{Portion of Galaxy Pair}  &  \colhead{$\sigma_v$\tablenotemark{a}}  &
  \colhead{corrected $\sigma_v$\tablenotemark{a}}\\
                                &   \colhead {(\kms)}  & \colhead{(\kms)}\\
 }
\startdata
Spiral arms of NGC 2207\tablenotemark{b}     &\\
(W1) west of central bar             &  $12 \pm 11$     &  $10 \pm 10$ \\
(W2) east of central bar              &  $18 \pm 17$     &  $16 \pm 16$ \\
&\\
Features in IC 2163\tablenotemark{b}           &\\
(W3) northern eyelid                 & $42 \pm 21$      &  $ 37 \pm 20$ \\ 
(W4) IR 8 + IR 9\tablenotemark{c}   & $15 \pm 6$  &  $ 12 \pm 6$ \\
(W5) southern eyelid                 &   $36 \pm 30$       &  $ 34 \pm 30$ \\                                      
\enddata 
\tablenotetext{a} {mean of unblanked pixels $\pm$ rms scatter in the region}
\tablenotetext{b} {see boxes marked in  Figure\,\ref{fig8} }.
\tablenotetext{c} {IR 8 and IR 9 are the IRAC clumps marked in Figure\,\ref{fig9}.}
\end{deluxetable}  

The CO velocity dispersion in a subtantial part of the eyelids is 
significantly higher than on the spiral arms of NGC 2207. 
As discussed below, we 
exclude from the eyelids the three dynamically complex regions 
labelled A, B, and C in Figure\,\ref{fig8}; they lie outside the range of $\theta$ studied
in Table\,\ref{table2}. Over much of the eyelids, the values of  $(\sigma_v)_{\rm corr}$ are high.
 However both eyelids have some regions, e.g., the
IRAC clumps IR 8 plus IR 9 on the northern eyelid and IR 3 on the southern eyelid
(see Figure\,\ref{fig9} for the locations of these clumps), 
where $(\sigma_v)_{\rm corr}$ is nearly the same as \citet{mogotsi16} find
in normal  galaxies. Our method of 
correcting  $\sigma _v$ for the  mean velocity gradient across the synthesized beam
introduced considerable blanking at these locations. The velocity dispersion 
before correcting for the velocity gradient (see the top panel of  Figure\,\ref{fig8} and
Table\,\ref{dispersion}) is also not particularly high in these regions. 

   We rule out stellar feedback as the source of the high values of
$(\sigma_v)_{\rm corr}$ on the eyelids by
 comparing the locations of the bright clumps of 24 \micron\ and \Halpha\ emission in
IC 2163 (see Figure\,\ref{fig9}) with where the high values of velocity dispersion occur.
 The northern eyelid clumps IR 11, IR 10, IR 9, and IR 8 and the
 southern eyelid clumps IR 5, IR 4, and IR 3 are the most luminous 
clumps of 24 \micron\ emission in IC 2163 and are also bright in \Halpha.
Of these, IR 11, IR 10, IR 5 (which partially overlaps Region B), and IR 4 
have $(\sigma_v)_{\rm corr} \geq 34$ \kms, whereas IR 8, IR 9, and IR 3 have 
$(\sigma_v)_{\rm corr} \simeq$ 11 \kms. 
IR 9 is the second most luminous star-forming region in NGC 2207/IC 2163 in
8 \micron, 24 \micron, and $\lambda 6$ cm radio continuum emission \citep{elmegreen06,
kaufman12, elmegreen16}.   
In contrast, on the northern eyelid, the locations labelled C1 and C2 
in Figure\,\ref{fig9} 
have  $(\sigma_v)_{\rm corr}$ = 27 \kms\ and 35 \kms, resp.,  but do not
contain bright clumps of 24 \micron\ or \Halpha\ emission.  

Turbulence generated by the ocular shock front remains as a
 viable explanation for the high values of  $(\sigma_v)_{\rm corr}$ on much of the eyelids.
The observed velocity dispersions may be an immediate result of the impact and shear
 in the eyelids or a subsequent development of microturbulence (random motions 
on small scales) as the components of the gas mix. Gravitational energy powers the
microturbulence via gravito-hydrodynamic instabilities. The regions of  
low $(\sigma_v)_{\rm corr}$ on both eyelids may be where the microturbulence
dissipated, the energy was radiated away, and the gas was not stirred up again.
These regions are farthest along the northern eyelid from the tidal bridge arm  and
somewhat far along the southern eyelid from the tidal tail.
  The presence of high velocity dispersion
on substantial portions of the eyelids seems consistent with the prolonged effect of
the prograde tidal force exerted by NGC 2207 on IC 2163 and thus in accord  with the
numerical models for the encounter by \citet{elmegreen95b, elmegreen00} and
\citet{struck05}.

Gas streaming radially inward from the 
outer part of the galaxy may dissipate a lot of translational 
kinetic energy via shock heating or microturbulence in the
eyelid compression zone. The amount of heating would depend on the details of how
the gas flows into the eyelids, which is beyond what we can derive from the present data.
\citet{kaufman12} find that the flux density ratio of 8 \micron\ to 
\sixcm\ radio continuum emission is a factor of two greater in the IRAC clumps on
the eyelids than elsewhere in this galaxy pair, possibly an indication of  
shock-heated H$_2$ in the eyelids.

\begin{figure*}
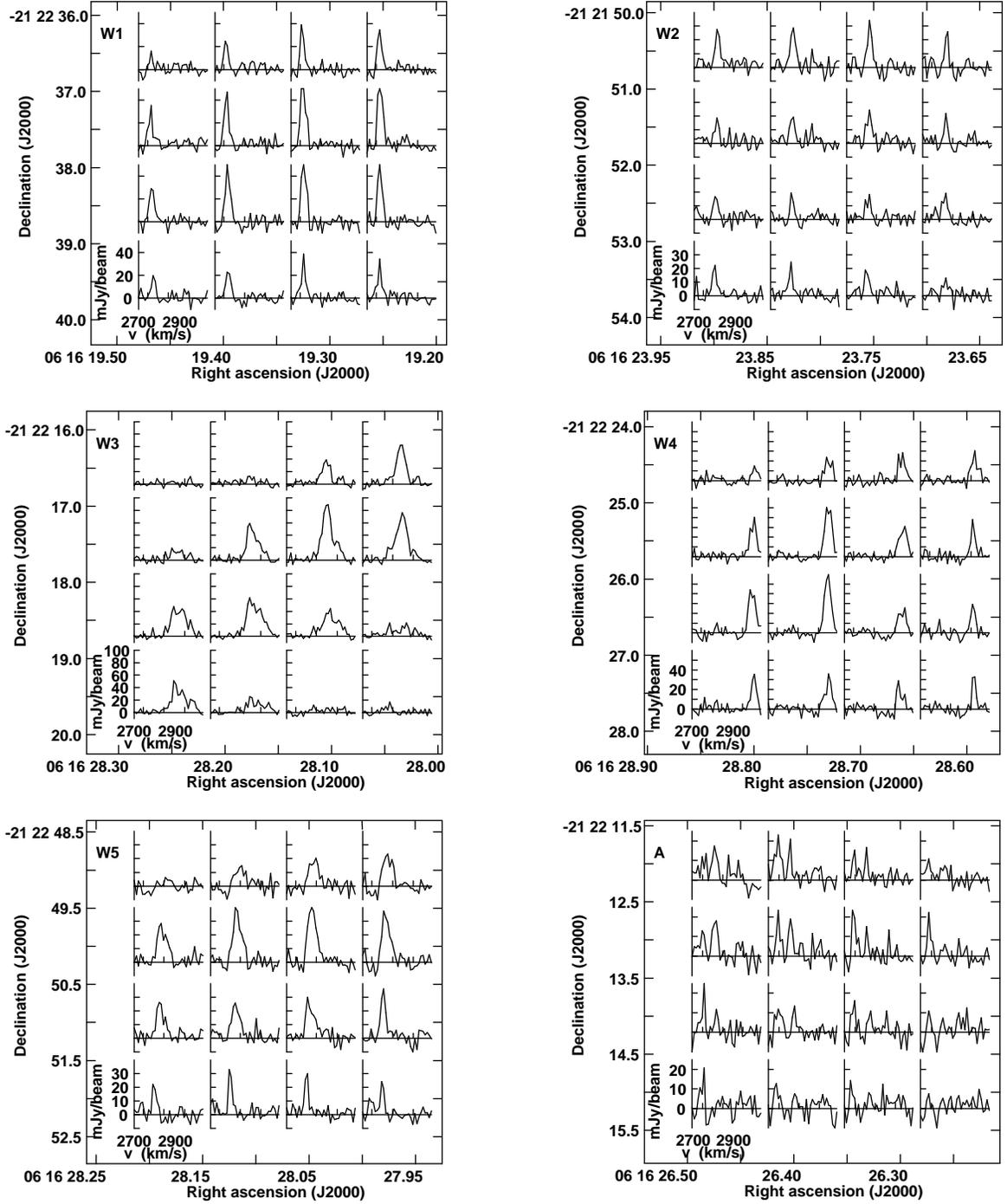

\epsscale{1.0}
\plottwo{figure10a.eps}{figure10b.eps}
\plottwo{figure10c.eps}{figure10d.eps}
\plottwo{figure10e.eps}{figure10f.eps}
\caption{Samples of CO line profiles, spaced $1''$ apart, from the unmasked cube for
 the various boxes
marked with W's in Figure\,\ref{fig8} and from Region A. Each set of line profiles has the
same velocity scale but, as labelled, may differ in intensity scale. The rms noise is 3.7 \mJybeam.
On the eyelids, these samples are from the locations of IR 10, IR 9, and IR 135
(the Spitzer clumps labelled in Figure\,\ref{fig9}).  
The high values of  $(\sigma_v)_{\rm corr}$ in Region A result from separate  
features, not high turbulence. 
\label{fig10}}
\end{figure*}

Figure\,\ref{fig10} displays samples of line profiles from the  unmasked cube
for the various boxes marked with W's in Figure\,\ref{fig8} and from Region A. 
At a number of positions in these W-box samples, we compare
the value of $\sigma_v$ from fitting the line profile in the unmasked cube 
with a single Gaussian to the value of $\sigma_v$ from the 
 second moment image 
in the top panel of Figure\,\ref{fig8}  (i.e., from the masked cube).
The ratio of the
Gaussian $\sigma _v$ to the second moment $\sigma_v$ has a mean value 
of 0.98, with an rms scatter of $\pm 0.18$.
In
box W3 and much of box W5 on the eyelids, the line profiles are broad; the high values of 
$(\sigma_v)_{\rm corr}$ result here from elevated turbulence. The line profiles in boxes
W1 and W2 in NGC 2207 are narrow. 

Region A  at $\theta \simeq 240\degr$ is where the northern eyelid joins the tidal bridge arm, 
and Region C at $\theta \simeq 60\degr$ is where the southern eyelid joins the tidal tail. 
These are regions of complicated  dynamics where the orbits change [see, for example, Fig. 2 in
\citet{elmegreen00}]. Region C is part of the massive \HI\ cloud I3 (see Section 3.2),
which has a mean \HI\ velocity dispersion of 51 \kms.
 Region B ($\theta \simeq 180\degr$) where the inner spiral
arm seems to join the southern eyelid   is another region of complex dynamics.
In Region A, many of the line profiles consist of two narrow peaks separated by
$> 60$ \kms, and thus Region A is not a region of high turbulence. In Regions B and C, the 
situation from the line profiles is less clear.
 
\section{Conclusions}

In the encounter simulations by \citet{elmegreen95b, elmegreen00} and 
\citet{struck05} for this galaxy pair, the  prograde tidal force exerted on IC 2163 
pulls outward on two opposite quadrants, producing the tidal tail and the tidal bridge,
 and pinches inward on the other two quadrants, producing the ocular structure. 
Although close, prograde, approximately in-plane collisions between disk  galaxies
 of similar mass are not uncommon, only a few galaxies exhibiting ocular structure 
are known  \citep{kaufman97, kaufman99} because the ocular stage lasts 
for only an short time, e.g., the duration of the present ocular disturbance in IC 2163
is of the order of several $\times 10^7$ yr.
Our ALMA \CO\ observations of the galaxy pair NGC 2207/IC 2163
with $2'' \times 1.5''$ resolution provide the first observational confirmation 
from velocity data with sufficiently high spatial resolution of how ocular structure
develops in such an encounter. This supports 
numerical models for the effects of close, prograde encounters in general.

We find large LOS velocity differences $|\Delta v|$ across the $\sim 6''$ 
(= 1 kpc) width of each eyelid of the ocular, 
some in excess of 100 \kms\ over a wide range of azimuths. These are a mixture of
radial and azimuthal streaming of gas at the outer edge of the eyelid relative to
its inner edge. The sense of the radial streaming at the eyelids is
consistent with the idea that material from the outer part of IC 2163 flows inward  
 until it hits an angular momentum barrier, where the radial streaming slows down 
abruptly 
and produces the observed galactic-scale pileup/shock zone at the eyelids. 
Across the 1 kpc width of each eyelid, the corresponding 
 values of  $|dv/dR|$ in the disk-plane are large, with some on the northern 
eyelid in excess of 150 \kms\ kpc$^{-1}$. 

We note velocity asymmetries between the  northern and southern eyelids and
conjecture that these may result from differences in the near-field tidal forces 
on the two eyelids.

Radial streaming into the eyelids causes an increase in the gas column density by
direct radial impact and also, with angular momentum conservation,
 leads to a high rate of shear.
We find a strong correlation at fixed values of azimuth between the 
molecular column density and $|dv/dR|$. Such a correlation may arise from a
combination of the shocks produced by direct radial impact and the shocks in
the intersecting gas streams resulting from shear.

Substantial portions of the eyelids have high CO velocity dispersion, which is
not  a product of stellar feedback. Since turbulence generated by the ocular shock front
is an expected consequence of the prolonged 
prograde tidal force of  NGC 2207 on IC 2163, 
the presence of high velocity  dispersion on the eyelids is
 in accord  with the above numerical models for the encounter.

The SPH model for NGC 2207/IC 2163 by \citet{struck05} has  a full treatment of both galaxies
simultaneously and handles gas plus stars but does not have sufficient number 
of particles or spatial resolution at the level of the eyelid width
for predicting the values of $\Delta v$ as a function of azimuth in the eyelids.  The latter is left
for future models.

\acknowledgments

   This paper makes use of the following ALMA data:
   ADS/JAO.ALMA\#2012.1.00357.S. ALMA is a partnership of ESO (representing
   its member states), NSF (USA) and NINS (Japan), together with NRC
   (Canada) and NSC and ASIAA (Taiwan) and KASI (Republic of Korea), in
   cooperation with the Republic of Chile. The Joint ALMA Observatory is
   operated by ESO, AUI/NRAO and NAOJ.
  The National Radio Astronomy Observatory is a facility of the National
  Science Foundation operated under cooperative agreement by Associated
   Universities, Inc. This research made use of the NASA/IPAC Extragalactic 
 Database (NED) which is operated by the Jet Propulsion Laboratory, California 
Institute of Technology, under contract with the National Aeronautics and Space
Administration. EB acknowledges support from the UK Science and Technology
Facilities Council [grant number ST/M001008/1].  FB acknowledges funding 
from the EU through grant ERC-StG-257720.

We thank Thangasamy Velusamy for providing their HiRes 24 \micron\ FITS image
to us. We thank the referee for making detailed comments and constructive suggestions.



\clearpage




\clearpage

\begin{thebibliography}{}
\bibitem[Dame et al.(2001)] {dame01} Dame, T.M., Hartmann, D., \& Thaddeus, P.  2001, 
  \apj, 655, 863

\bibitem[Donner et al.(1991)] {donner91} Donner, K.J, Engstrom, S. \& Sundelius, B. 1991, 
  A\&A, 252, 571

\bibitem[Elmegreen et al.(1991)] {elmegreen91} Elmegreen, D.M., Sundin, M.,
    Elmegreen, B.G., \& Sundelius, B. 1991, A\&A, 244, 52

\bibitem[Elmegreen et al.(1995a)] {elmegreen95a} Elmegreen, D.M., 
 Kaufman, M., Brinks, E., Elmegreen, B. G., \& Sundin, M. 1995a,
  \apj,  453, 100

\bibitem[Elmegreen et al.(1995b)] {elmegreen95b} Elmegreen, B.G.,  Sundin, M.,
Kaufman, M., Brinks, E., \& Elmegreen, D.M. 1995b, \apj,  453, 139

\bibitem[Elmegreen et al.(1998)] {elmegreen98} Elmegreen, B.G., Elmegreen, D.M.,
Brinks, E., et al. 1998, \apj, 503, L119

\bibitem[Elmegreen et al.(2000)] {elmegreen00} Elmegreen, B.G., Kaufman, M., 
Struck, C., et al. 2000, \aj, 120,  630.

\bibitem[Elmegreen et al.(2001)] {elmegreen01} Elmegreen, D.M., Kaufman, M.,
Elmegreen, B.G.,  Brinks, E., Struck, C., Klaric, M., \& Thomasson, M. 2001, 
\aj, 121, 182

\bibitem[Elmegreen et al.(2006)] {elmegreen06} Elmegreen, D.M., Elmegreen, B.G., 
Kaufman, M., Sheth, K., Struck, C., Thomasson, M., \& Brinks, E.  2006,  \apj,  
642, 158

\bibitem[Elmegreen et al.(2016)] {elmegreen16} Elmegreen, B.G., Kaufman, M., Bournaud, F.,
Elmegreen, D.M., Struck, C., Brinks, E., \& Juneau, S.  2016,  \apj, 823, 26

\bibitem[Kaufman et al.(1997)] {kaufman97} Kaufman, M., Brinks, E., Elmegreen, D.M,
     Thomasson, M., Elmegreen, B.G., Struck, C., \& Klari\'{c}, M.  1997,  \aj, 114, 2323

\bibitem[Kaufman et al.(1999)]{kaufman99} Kaufman, M., Brinks, E., Elmegreen, B.G.,
     Elmegreen, D.M.,  Klari\'{c}, M., Struck, C., Thomasson, M., \& Vogel, S.  1999, 
     \aj, 118, 1577

\bibitem[Kaufman et al.(2012)] {kaufman12} Kaufman, M., Grupe, D., Elmegreen, B.G.,
Elmegreen, D.M., Struck, C., \& Brinks, E. 2012, \aj, 144, 156

\bibitem[Mineo et al.(2014)]{mineo14} Mineo, S., Rapport, S., Levine, A., Pooley, D., 
Steinhorn, B., \&  Homan, J.  2014, \apj, 797, 91 

\bibitem[Mogotsi et al.(2016)]{mogotsi16} Mogotsi, K.M.,  de Blok, W.J.G., Cald\'{u}-Primo, A.,
     Walter, F., Ianjamasimanana, R., Leroy, A.K., \& Leroy, A.K. 2016,  \aj, 151, 15

\bibitem[Struck et al.(2005)]{struck05} Struck, C., Kaufman, M., Brinks, E.,
Thomasson, M., Elmegreen, B.G., \& Elmegreen, D.M. 2005, \mnras, 364, 69

\bibitem[Sundin(1989)] {sundin89} Sundin, M. 1989,  in ''Dynamics of Astrophysical
Disks," ed. J. Sellwood, Cambridge University Press, p. 215

\bibitem[Thomasson(2004)]{thomasson04} Thomasson, M.  2004,  in ASP Conf. Ser. 320,
    The Neutral ISM in Starburst Galaxies, ed. S. Aalto, S.  H\"{u}ttemeister, \& A. Pedlar 
    (San Francisco, CA: ASP), 81

\bibitem[Velusamy et al.(2008)]{velusamy08} Velusamy, T., Marsh, K.A., Beichman, C.A.,
    Backus, C.R., \& Thompson, T.J.  2008, \aj, 136, 197

\end{thebibliography}
\end{document}